\documentclass[letterpaper]{IEEEtran}
\usepackage{graphicx}
\usepackage{physics}
\usepackage{amsmath}
\usepackage{amssymb}
\usepackage{enumerate}
\usepackage{framed}
\usepackage{adjustbox}
\usepackage{multirow}
\usepackage{multicol}
\usepackage{listings}
\usepackage{tikz}
\usepackage{xcolor}
\usepackage{bm}
\usepackage{mathtools}
\usepackage{algorithm}
\usepackage{algpseudocode,eqparbox}
\usepackage{caption}

\mathtoolsset{showonlyrefs}
\allowdisplaybreaks
\usepackage{letltxmacro}
\LetLtxMacro{\originaleqref}{\eqref}
\renewcommand{\eqref}{Eq.~\originaleqref}

\definecolor{codegreen}{rgb}{0,0.6,0}
\definecolor{codegray}{rgb}{0.5,0.5,0.5}
\definecolor{codepurple}{rgb}{0.58,0,0.82}
\definecolor{backcolour}{rgb}{0.95,0.95,0.92}

\lstdefinestyle{mystyle}{
    backgroundcolor=\color{backcolour},   
    commentstyle=\color{codegreen},
    keywordstyle=\color{magenta},
    numberstyle=\tiny\color{codegray},
    stringstyle=\color{codepurple},
    basicstyle=\ttfamily\footnotesize,
    breakatwhitespace=false,         
    breaklines=true,                 
    captionpos=b,                    
    keepspaces=true,                 
    numbers=left,                    
    numbersep=5pt,                  
    showspaces=false,                
    showstringspaces=false,
    showtabs=false,                  
    tabsize=2
}

\usepackage{stmaryrd}

\usepackage[style=ieee,backend=biber]{biblatex}
\bibliography{isit}

\usepackage{ifthen}

\newif\ifarxiv
\arxivtrue

\makeatletter
\newcounter{IEEE@bibentries}
\renewcommand\IEEEtriggeratref[1]{%
  \renewbibmacro{finentry}{%
    \stepcounter{IEEE@bibentries}%
    \ifthenelse{\equal{\value{IEEE@bibentries}}{#1}}
    {\finentry\@IEEEtriggercmd}
    {\finentry}%
  }%
}
\makeatother

\usepackage{mathtools}

\usetikzlibrary{shapes.geometric, arrows}
\tikzstyle{startstop} = [rectangle,  minimum width=3cm, minimum height=1cm,text centered,text width=10cm, draw=black ,fill=gray!20]
\tikzstyle{process} = [rectangle, minimum width=3cm, minimum height=1cm, text centered,text width=10cm, draw=black,fill=orange!20]
\tikzstyle{arrow} = [thick,->,>=stealth]
\tikzstyle{state}=[shape=circle,draw=blue!50,fill=blue!20]
\tikzstyle{observation}=[shape=rectangle,draw=orange!50,fill=orange!20]
\tikzstyle{lightedge}=[<-,dotted]
\tikzstyle{mainstate}=[state,thick]
\tikzstyle{mainedge}=[<-,thick]
\usepackage{color}
\definecolor{bitcolor}{rgb}{1,0.84314,0}
\definecolor{checkcolor}{rgb}{0.52941,0.80784,1}
\usepackage{pgfplots}
\pgfplotsset{compat=1.18}
\renewcommand{\epsilon}{\varepsilon}

\newcommand{\vnop}{\varoast}

\newcommand{\cnop}{\boxast}

\newtheorem{theorem}{Theorem}

\newtheorem{rem}[theorem]{Remark}
\newtheorem{lem}[theorem]{Lemma}
\newtheorem{defn}[theorem]{Definition}

\date{}

\title{Polar Codes for CQ Channels: Decoding via Belief-Propagation with Quantum Messages}

\IEEEoverridecommandlockouts
\author{%
   \IEEEauthorblockN{Avijit Mandal\IEEEauthorrefmark{1}\IEEEauthorrefmark{2}, Sam Brandsen\IEEEauthorrefmark{4}, and Henry D. Pfister\IEEEauthorrefmark{1}\IEEEauthorrefmark{2}\IEEEauthorrefmark{3} \thanks{This research was supported in part by the National Science Foundation (NSF) under Grants 1908730, 2106213, and 2212437. Any opinions, findings, and conclusions or recommendations expressed in this material are those of the author(s) and do not necessarily reflect the views of the NSF.}} \\
   \IEEEauthorblockA{%
                  Duke Quantum Center, Duke University, 
                 Durham, NC, USA\IEEEauthorrefmark{1}  } \\
    \IEEEauthorblockA{Dept.\ of Electrical and Computer Engineering, Duke University, Durham, NC, USA\IEEEauthorrefmark{2}} \\
   \IEEEauthorblockA{Dept.\ of Mathematics, Duke University, Durham, NC, USA\IEEEauthorrefmark{3}} \\
   \IEEEauthorblockA{Dept. of Psychiatry and Behavioral Sciences, Duke University, Durham, NC, USA\IEEEauthorrefmark{4}} }

\begin{document}

\maketitle
\begin{abstract}
This paper considers the design and decoding of polar codes for general classical-quantum (CQ) channels.
It focuses on decoding via belief-propagation with quantum messages (BPQM) and, in particular, the idea of paired-measurement BPQM (PM-BPQM) decoding.
Since the PM-BPQM decoder admits a classical density evolution (DE) analysis, one can use DE to design a polar code for any CQ channel and then efficiently compute the trade-off between code rate and error probability.
We have also implemented and tested a classical simulation of our PM-BPQM decoder for polar codes.
While the decoder can be implemented efficiently on a quantum computer, simulating the decoder on a classical computer actually has exponential complexity.
Thus, simulation results for the decoder are somewhat limited and are included primarily to validate our theoretical results.
\end{abstract}


\section{Introduction}
The study of channel coding for CQ channels dates back to Holevo, Schumacher, and Westmoreland \cite{holevo1998capacity} and \cite{schumacher1997sending}.
For a comprehensive introduction, see \cite{nielsen2002quantum,wilde2013quantum}.
This paper provides a detailed description of how to efficiently design (on a classical computer) and decode (on a quantum computer) polar codes for classical-quantum (CQ) 
channels.
Practical applications motivating coding for CQ channels can be found in \cite{guha2012polar,krovi2015optimal,da2013achieving}.

Polar codes for CQ channels were first introduced in 2012~\cite{wilde2012polar}.
A variety of follow-up papers were able to extend and improve these results~\cite{guha2012polar,wilde2013towards,renes2014polar, nasser2018polar}.
Although these papers describe a design and decoding process for polar codes on CQ channels in theory, efficient algorithms (e.g., polynomial time in the block length) are not described for either of these tasks.

In 2017, Renes describes a belief-propagation with quantum messages (BPQM) algorithm that provides an optimal decoding method for binary linear codes with tree factor graphs on the pure-state channel (PSC)~\cite{Renes-njp17}.
That work notes that BPQM could allow efficient decoding of polar codes on the PSC.
BPQM is explored further in~\cite{rengaswamy2021belief} and made more efficient by Renes and Piveteau in \cite{piveteau2022quantum}.
Recently, an extension of BPQM for general CQ channels was introduced and called paired-measurement BPQM (PM-BPQM)~\cite{brandsen2022belief}.
This algorithm can be applied to any symmetric binary-input CQ channel and is equipped with a classical DE analysis of its decoding performance.
In \cite{brandsen2022belief}, this DE analysis is used to compute noise thresholds for PM-BPQM decoding of low-density parity-check (LDPC) codes \cite{gallager1962low} on CQ channels.

In this paper, we use DE to design polar codes for PM-BPQM decoding and determine their achievable rates.
We also report results for the classical simulation of our PM-BPQM decoder for polar codes.
Although this decoder can be implemented efficiently on a quantum computer, directly simulating the decoder on a classical computer has exponential complexity in the block length.
Thus, our simulation results are limited to very short block lengths and are included mainly to validate our DE results.

\section{Background}

\subsection{Preliminaries}
We define the set of natural numbers by $\mathbb{N}=\left\{ 1,2,\ldots\right\} $ and use the shorthand $[m]\coloneqq\left\{ 1,\ldots,m\right\} $ for $m\in\mathbb{N}$.
Let $\mathcal{H}_{n}$ denote the $n$-dimensional Hilbert space $\mathbb{C}^n$.
A quantum \emph{pure state} is a unit length vector $\ket{\psi}\in \mathcal{H}_{n}$. For  quantum systems $A_{1},A_{2} \dots ,A_{n}$, we denote the joint quantum state of $n$ qubits, $\ket{\psi}_{A_{1}A_{2}\dots A_{n}}$ where we follow the convention that the $1$\textsuperscript{st} qubit is associated with the system $A_{1}$, the $2$\textsuperscript{nd} qubit with $A_{2}$ and so on. When $\ket{\psi}_{A_{1}A_{2}\dots A_{n}}$ is not entangled, we can use the Kronecker product to write the joint state as 
\begin{align}
    \ket{\psi}_{A_{1}A_{2}\dots A_{n}}=\ket{\psi_{1}}_{A_{1}}\otimes\dots\ket{\psi_{n}}_{A_{n}}.
\end{align}
A random ensemble of $m$ quantum pure states in $\mathcal{H}_n$ is denoted by $\Psi = \{ p_{i},\ket{\psi_{i}} \} |_{i=1}^m$, where $p_{i}$ denotes the probability of choosing the pure state $\ket{\psi_{i}}$. 
This ensemble can also be represented by the \emph{density matrix}
$\rho=\sum_{i=1}^m p_{i}\ketbra{\psi_{i}}{\psi_{i}}\in \mathbb{C}^{n\times n}$.
All such density matrices are positive semidefinite with unit trace and we use $\mathcal{D}(\mathcal{H}_{n})$ to denote this subset.
The unitary evolution of a quantum state $\ket{\psi} \in \mathcal{H}_n$ is described by the mapping $\ket{\psi} \mapsto U \ket{\psi}$, where $U\in \mathbb{C}^{n\times n}$ is a unitary.
For the pure state ensemble $\Psi$, this evolution results in the modified ensemble $\Psi'=\{p_{i},U\ket{\psi_{i}}\} |_{i=1}^m$ whose density matrix is \vspace{-1.5mm}
\begin{align*}
    \rho' = \sum_{i=1}^m p_{i}U\ketbra{\psi_{i}}{\psi_{i}}U^{\dagger}=U\rho U^{\dagger},
\end{align*}
where $U^\dagger$ is the Hermitian transpose of $U$. We denote the Pauli matrices by
\begin{align*}
\sigma_{x} \coloneqq \begin{bmatrix}0 & 1 \\ 1& 0 \end{bmatrix}, \sigma_{y} \coloneqq \begin{bmatrix}0 & -i \\ i& 0 \end{bmatrix}, \sigma_{z} \coloneqq \begin{bmatrix} 1& 0 \\ 0 & -1 \end{bmatrix}.
\end{align*}

\begin{defn}
A binary symmetric CQ (BSCQ) channel is defined by the mapping $W:\{0,1\}\rightarrow \mathcal{D}(\mathcal{H}_{n})$ from the binary classical input $z\in\{0,1\}$ to the density matrix $W(z)\in\mathcal{D}(\mathcal{H}_{n}) $ of the quantum output and a unitary $U$ satisfying $U^{2}=\mathbb{I}$.
The unitary defines the symmetry constraint via  $W(1)=UW(0)U^\dagger$. 
\end{defn}
\begin{lem}[{\cite[Appendix~A]{brandsen_bpqm_arxiv}}]
\label{lem:symmetricCQ}
Any BSCQ channel that outputs a qubit is unitarily equivalent to the qubit channel $W:\left\{ 0,1\right\} \to\mathcal{D}(\mathcal{H}_{2})$ satisfying $W(z)=\sigma_{x}^{z}\rho(\delta,\gamma)\sigma_{x}^{z}$ with \vspace*{-1mm}
\[
\rho(\delta,\gamma)\coloneqq \begin{bmatrix}\begin{array}{cc}
\delta & \gamma^{*}\\
\gamma & 1-\delta
\end{array}\end{bmatrix} \vspace*{-1mm}
\]
for some $\delta\in[0,1]$ and $\gamma\in\mathbb{C}$ satisfying $\left|\gamma\right|^{2}\leq\delta(1-\delta)$.

 \end{lem}

This representation enables one to characterize any qubit BSCQ channel using only two parameters $\delta$ and $\gamma$.

\begin{defn}
An $m$-outcome \emph{projective measurement} of a quantum system in $\mathcal{H}_n$ is defined by a set of $m$ orthogonal projection matrices $\Pi_{j} \in \mathbb{C}^{n\times n}$ satisfying $\Pi_{i}\Pi_{j}=\delta_{i,j}\Pi_{i}$ and $\sum_{j}\Pi_{j}=\mathbb{I}_{n}$, where $\mathbb{I}_n$ is the $n\times n$ identity matrix.
We denote such a measurement by $\hat{\Pi}=\{\Pi_{j}\}|_{j=1}^{m}$.
\end{defn}

Applying the measurement $\hat{\Pi}$ to the quantum state $\rho$ results in a random outcome $J$ and the probability of the event $J=j$ is given by $\text{Tr}(\Pi_{j}\rho)$.
The post-measurement state, conditioned on the event $J=j$, is given by $\Pi_{j}\rho\Pi_{j}/ \text{Tr}(\Pi_{j}\rho)$.

Consider a hypothesis test to distinguish between $m$ possible quantum states defined by $ \Phi = \{p_j,\rho_{j}\} |_{j=1}^{m}$, where the $j$-th hypothesis has prior probability $p_j$ and density matrix $\rho_j$.
For a projective measurement $\hat{\Pi}$, where $\Pi_j$ is associated with hypothesis $\rho_j$, the probability of choosing correctly is \vspace{-1mm}
\begin{align*}
    P(\Phi,\hat{\Pi}) =\sum_{j=1}^m p_{j}\text{Tr}(\Pi_{j}\rho_{j}). \vspace*{-1mm}
\end{align*}
\begin{defn}
The \emph{Helstrom measurement} is the minimum-error  measurement to distinguish between two density matrices $\rho_{0},\rho_{1} \in \mathcal{D}(\mathcal{H}_n)$ when $\rho_0$ has prior  probability $p$.
This measurement maximizes the success probability of the test by forming projection operators onto the positive and negative eigenspaces of $M=p\rho_{0}-(1-p)\rho_{1}$.
Formally, it is defined by $\hat{\Pi}_{H}=\{\Pi_{+},\mathbb{I}_{n}-\Pi_{+}\}$, where \vspace*{-1mm}
\begin{align*}
    \Pi_{+}=\sum_{\ket{v}\in \mathcal{V}_{+}}\ketbra{v}{v} \vspace*{-1mm}
\end{align*}
where, $\mathcal{V}_{+}=\{\ket{v} \in \mathcal{H}_n \,|\, \braket{v}{v}=1,\exists \lambda\geq 0, M\ket{v}=\lambda \ket{v}\}$.
\end{defn}

\begin{lem}[{\cite[Lemma~6]{brandsen2022belief}}]
Consider a a BSCQ channel $W:\{0,1\}\rightarrow \mathcal{D}(\mathcal{H}_{2n})$ with equiprobable outputs where $W(z)=U^{z}\rho U^{z}$.
Then, the Helstrom measurement $\hat{\Pi}_{H}$ is defined by \vspace*{-1mm}
\begin{align}
   \hat{\Pi}_{H}=\left\{\sum_{j=0}^{n-1}\ketbra{v_{j}}{v_{j}},\sum_{j=0}^{n-1}U\ketbra{v_{j}}{v_{j}}U\right\}, \vspace*{-1mm}
\end{align}
where $\{\ket{v_{j}}\}_{j=0}^{n-1}$ is the set of eigenvectors for $W(0)-W(1)$ with non-negative eigenvalues.
\end{lem}

\begin{rem} \label{rem:qbscq_err}
For a qubit BSCQ with parameters $(\delta,\gamma)$, the error rate of the Helstrom measurement is $\delta$.  
\end{rem}

\subsection{Paired-Measurement BPQM}

Paired-measurement BPQM (PM-BPQM) was introduced in~\cite{brandsen2022belief} as a generalization of BPQM~\cite{Renes-njp17}.
\begin{lem}[{\cite[Lemma~6]{brandsen2022belief}}]
Consider the Helstrom measurement to distinguish between $W(0)$ and $W(1)$ for the BSCQ channel $W:\{0,1\}\rightarrow \mathcal{D}(\mathcal{H}_{n})$.
One can achieve the same error rate by first implementing
\[\hat{\Pi}'=\Big{\{}\ketbra{v_{j}}{v_{j}}+U\ketbra{v_{j}}{v_{j}}U\Big{\}}_{j=0}^{n-1}\]
and then, if the first outcome is $j$, implementing
\[\hat{\Pi}(j)=\Big{\{}\ketbra{v_{j}}{v_{j}},U\ketbra{v_{j}}{v_{j}}U\Big{\}}.\]
\end{lem}
\begin{lem}[{\cite[Lemma~7]{brandsen2022belief}}]\label{bscq_post measurement update}
Consider a BSCQ channel $W: \{0, 1\} \rightarrow \mathcal{D}(\mathcal{H}_{n})$ with $W(0)=\rho$ and equiprobable inputs.
Then, this channel followed by the paired measurement $\hat{\Pi}_{W}$ gives a new CQ channel which is a distinguishable mixture of symmetric qubit channels defined by
\[
\widetilde{W}(z) \mapsto\sum_{j=0}^{m-1}p_{j}\Big(\sigma_{x}^{z}\rho\sigma_{x}^{z}\otimes\ket{j}\!\!\bra{j} \Big),
\]
where the $j$-th paired outcome has probability $p_j = \Tr\big[\big(\ketbra{v_{j}}{v_{j}} + U \ketbra{v_{j}}{v_{j}} U\big) \rho \big]$ and post-measurement density matrix
\[ \rho_j = \frac{1}{p_j} \begin{pmatrix}
    \bra{v_{j}} \rho \ket{v_{j}} & \bra{v_{j}} U \rho \ket{v_{j}} \\
    \bra{v_{j}} \rho U \ket{v_{j}} & \bra{v_{j}} U \rho U \ket{v_{j}}
    \end{pmatrix}. \]
\end{lem}
From Lemma~\ref{lem:symmetricCQ}, we can identify the parameters of $\rho_{j}$ via
\begin{equation}
    \delta_{j}
     = \frac{1}{p_{j}}\bra{v_{j}}\rho\ket{v_{j}}, \quad\quad
    \gamma_{j}=\frac{1}{p_{j}}\bra{v_{j}}U\rho\ket{v_{j}}.
\end{equation}

Now, we will describe the channel combining operations that are used to define the PM-BPQM updates~\cite{Renes-njp17,brandsen2022belief}.
For binary CQ channels $W,W'$, the check-node and bit-node channel combining operations are defined by
\begin{align}
[W\cnop W'](z) & \coloneqq\frac{1}{2}\sum_{z'\in\{0,1\}}W(z\oplus z')\otimes W'(z') \label{eq:cnop} \\
[W\vnop W'](z) & \coloneqq W(z)\otimes W'(z) \label{eq:vnop}.
\end{align}

\ifarxiv
The following Lemma is proved in Appendix~\ref{deffered proofs}.
\fi
\begin{lem} \label{lem:bscq-checknode}
For qubit BSCQ channels $W$ and $W'$, using the paired measurement to distinguish between $[W\cnop W'](0)$ and $[W\cnop W'](1)$ is equivalent to the unitary operation
\begin{align*}
    C=\frac{1}{\sqrt{2}}\begin{bmatrix}
     1 & 0 &0 & 1\\
     -1 & 0 &0 &1 \\
     0 & 1&1 &0\\
     0 &1& -1&0
    \end{bmatrix}
\end{align*}
to get $\tau \coloneqq C [W\cnop W'](z)C^{\dagger}$ followed by measurement of the second qubit.
This results in the state
\begin{align}\label{bscq checknode measurement}
\Pi_0 \tau \Pi_0 + \Pi_1 \tau \Pi_1 = \hspace{-3mm}\sum_{j\in\{0,1\}}p^{\cnop}_{j} \left( \sigma_{x}^{z}\rho(\delta_{j}^{\cnop},\gamma_{j}^{\cnop})\sigma_{x}^{z}\otimes \ketbra{j}{j} \right),\;\;
\end{align}
where $\Pi_{0}=\mathbb{I}_{2}\otimes \ketbra{0}{0}$ and $\Pi_1 = \mathbb{I}_4 - \Pi_{0}$.
If $W$ and $W'$ have the channel parameters $(\delta,\gamma)$ and $(\delta',\gamma')$, then we can define
$\ket{v_{0}}=\frac{1}{\sqrt{2}}(1,0,0,1)$ and $\ket{v_{1}}=\frac{1}{\sqrt{2}}(-1,0,0,1)$ to compute
\begin{align}
&\ p_{j}^{\cnop}(\delta,\gamma,\delta',\gamma')\\
&\ \quad\coloneqq \text{Tr}\big{(}(\ketbra{v_{j}}{v_{j}}+\sigma_{x}\otimes \mathbb{I}_{2}\ketbra{v_{j}}{v_{j}}\sigma_{x}\otimes \mathbb{I}_{2})(W\cnop W')[0]\big{)}, \\
 &\   \delta_{j}^{\cnop} (\delta,\gamma,\delta',\gamma')
    \coloneqq \frac{1}{p^{\cnop}_{j}} \bra{v_{j}}(W\cnop W')[0]\ket{v_{j}}, \label{d-checknode}\\
  &\  \gamma_{j}^{\cnop}(\delta,\gamma,\delta',\gamma') \coloneqq
    \frac{1}{p^{\cnop}_{j}} \bra{v_{j}}(\sigma_{x}\otimes \mathbb{I}_{2})(W\cnop W')[0]\ket{v_{j}}\label{g-checknode}.
\end{align}
\end{lem}

\begin{rem}
We call unitary $C$ the check-node unitary because it compresses the decision information from the check-node combining operation into the first qubit while keeping the reliability information in the second qubit. It is worth noting that the check-node unitary does not depend on the parameters of $W$ and $W'$.  
\end{rem}

\ifarxiv
The following Lemma is proved in Appendix~\ref{deffered proofs}.
\fi
\begin{lem}\label{bscq-bitnode}
For qubit BSCQ channels $W$ and $W'$ with parameters $(\delta,\gamma)$ and $(\delta',\gamma')$ respectively, implementing paired measurement to distinguish between $[W\vnop W'](0)$ and $[W\vnop W'](1)$ is equivalent to implementing the unitary $V = V(\delta,\gamma,\delta',\gamma')$ to get $\tau' \coloneqq V [W\otimes W'](z) V^{\dagger}$ followed by measurement of the second qubit.
Here, the rows of $V(\delta,\gamma,\delta',\gamma')$ are defined, from top to bottom, by $\ket{v_{0}'}$, $\ket{v_{1}'}$, $(\sigma_{x}\otimes \sigma_{x} ) \ket{v_{0}'}$ and $(\sigma_{x}\otimes \sigma_{x}) \ket{v_{1}'}$ in terms of the 
paired-measurement eigenvectors $\ket{v_{0}'}$ and $\ket{v_{1}'}$ that span the positive eigenspace of $(W\vnop W')[0]-(W\vnop W')[1]$. This results in the state \vspace{-1mm}
\begin{align}\label{bscq bitnode measurement}
\! \Pi_0 \tau' \Pi_0 + \Pi_1 \tau' \Pi_1 = \hspace{-3mm}\sum_{j\in \{0,1\}}p^{\vnop}_{j} \left(\sigma_{x}^{z}\rho(\delta_{j}^{\vnop},\gamma_{j}^{\vnop})\sigma_{x}^{z}\otimes \ketbra{j}{j} \right)\!, \;\;\;
\end{align}
where $\Pi_{0}=\mathbb{I}_{2}\otimes \ketbra{0}{0}$, $\Pi_1 = \mathbb{I}_4 - \Pi_{0}$, and we have \vspace{-1mm}
\begin{align}
    &\ p_{j}^{\vnop}(\delta,\gamma,\delta',\gamma')\\
    &\ \; \coloneqq \text{Tr}\big{(}(\ketbra{v_{j}'}{v_{j}'}+\sigma_{x}\otimes \sigma_{x}\ketbra{v_{j}'}{v_{j}'}\sigma_{x}\otimes \sigma_{x})(W\vnop W')[0]\big{)}\\
 &\   \delta_{j}^{\vnop} (\delta,\gamma,\delta',\gamma')
    \coloneqq \frac{1}{p^{\vnop}_{j}} \bra{v_{j}'}(W\vnop W')[0]\ket{v_{j}'}, \label{d-bitnode}\\
  &\  \gamma_{j}^{\cnop}(\delta,\gamma,\delta',\gamma') \coloneqq
    \frac{1}{p^{\vnop}_{j}} \bra{v_{j}'}(\sigma_{x}\otimes \sigma_{x})(W\vnop W')[0]\ket{v_{j}'}\label{g-bitnode}.
\end{align}
\end{lem}

\begin{rem}
 The unitary $V(\delta,\gamma,\delta',\gamma')$ is called the bit-node unitary. Similar to the check-node unitary, it compresses the decision information from the bit-node channel combining it into the first qubit while keeping the reliability information in the second qubit.
 Unlike the check-node unitary $C$, the eigenvectors $\ket{v_{0}'}$ and $\ket{v_{1}'}$ depend on the channel parameters and thus the unitary $V(\delta,\gamma,\delta',\gamma')$ does as well~\cite{brandsen2022belief}.
 \end{rem}
 
\subsection{Density Evolution for Paired-Measurement BPQM}

Density evolution (DE) is a tool widely used by coding theorists to analyze the asymptotic performance of BP decoding for long codes chosen from certain families~\cite{richardson2008modern}.
For BSCQ channels, the paired measurement~\cite{brandsen_bpqm_arxiv} compresses the decision information from check-node and bit-node combining into the first qubit while keeping reliability information in the second qubit.
By tracking how the message reliability evolves through this process, one can use DE to analyze the performance of PM-BPQM decoding for a code whose factor graph is a tree.
Applying DE to a long code, whose factor graph is a tree with sufficiently large depth, results in a threshold phenomenon that allows one to estimate the noise threshold (i.e., the maximum noise level where DE predicts successful decoding) for families of codes.
In~\cite{brandsen_bpqm_arxiv}, this was applied to regular LDPC codes on CQ channels with PM-BPQM decoding.

\subsection{Polar Codes}
Polar codes were introduced by Ar{\i}kan in 2009 as the first deterministic construction
of capacity-achieving codes for binary memoryless symmetric (BMS) channels~\cite{arikan2009channel}.
The polar transform of length $N=2^{n}$ is denoted by $G_{N}\triangleq B_{N}G_{2}^{\otimes n}$ where $B_{N}$ is $N\times N$  bit reversal matrix~\cite[Sec.~VII.B]{arikan2009channel} and $G_{2}^{\otimes n}$ is $n$-fold tensor product of $2\times 2$ binary matrix 
\begin{align*}
    G_{2}\triangleq\begin{bmatrix}
     1 & 0\\
     1 & 1
    \end{bmatrix}.
\end{align*}
Polar codes work by using the polar transform $G_N$ to encode a vector $u \in \{0,1\}^N$ whose values are free on a subset $\mathcal{A} \subseteq [N]$ of information positions but restricted to have fixed values on the complementary set $\mathcal{A}^c = [N] \setminus A$ of frozen positions.
For the frozen positions, the fixed values are shared with the receiver in advance to aid the decoding process.

Polar codes can achieve capacity on BMS channels under low-complexity successive-cancellation (SC) decoding, where one decodes the bits $u_1,u_2,\ldots$ in order assuming all past decoding decisions are correct but that no information is known about future $u$ values.
The performance of this approach is analyzed by recursively defining the effective channels seen by the SC decoder assuming all past decisions are correct.

Polar codes were extended to CQ channels by Wilde and Guha~\cite{wilde2012polar}.
Similar to the classical case, one can recursively define effective channels that characterize the performance of successive cancellation (SC) decoding.
When decoding the $i$-th bit of a length-$N$ polar code, the designed effective channel~\cite[p.~1178]{wilde2012polar} for a CQ channel $W$ is defined by
\[ W_N^{(i)} (u_i) \coloneqq \frac{1}{2^{N-1}} \!\!\!\!\!\!\!\! \sum_{u_{\sim i} \in \{0,1\}^{N-1}} \!\!\!\!\!\!\!\!\! \ketbra{u_1^{i-1}}{u_1^{i-1}} \otimes \left( \bigotimes_{i=1}^N W([\bm{u} G_N]_i)\! \right)\!. \]
The SC decoder implements the implied sequence of Helstrom measurements indexed by the information bits in $\mathcal{A}$.
One difference from the classical case is that the effective channels encountered during decoding may differ from the designed effective channels even when all past decisions are correct.
This is because the sequential measurement process can disturb the codeword state even when the decision is correct.
In particular, the channel seen by the decoder equals the designed effective channel if all earlier bits are frozen (i.e., $[i-1]\subseteq \mathcal{A}^c$).
When the set $\mathcal{A}$ of information channels is selected to achieve a sufficiently low error rate, the non-commutative union bound (e.g., see \cite{gao2015quantum,khabbazi2019union,o2022quantum}) shows that this disturbance is negligible. 

\section{PM-BPQM and Polar Codes}

\subsection{PM-BPQM DE for Polar Code Design}\label{sec:pmbpqm_polar_de}

To design a polar code for a CQ channel assuming SC decoding based on PM-BPQM,  we implement PM-BPQM DE for the BSCQ channel via Monte Carlo simulation.

Let $W_{N}^{(i)}$ refer to the effective channel experienced by the $i$-th information bit of a length-$N$ polar code assuming all previous bits are frozen.
The input-output law for this channel can be computed recursively using
\begin{align}
    W_{N}^{(2i-1)} &\ =W^{(i)}_{N/2}\cnop W^{(i)}_{N/2}\\
     W_{N}^{(2i)} &\ =W^{(i)}_{N/2}\vnop W^{(i)}_{N/2},
\end{align}
where the check-node and bit-node update rules are defined by~\eqref{eq:cnop} and~\eqref{eq:vnop}.
For a code of length $N=2^n$, this requires $n$ levels of recursion starting from $W=W_1^{(1)}$.

At each level of the recursion, one computes a representation of the new channels via check-node and bit-node updates from representations of the channels at the previous level.
This approach has two key issues.
First, the channels involved may not have simple representations.
While any classical BMS channel can be represented as a mixture of binary symmetric channels (i.e., a distribution over $[0,\frac{1}{2}]$)~\cite{richardson2008modern}, the set of BSCQ channels does have such a simple description.
Second, the set of possible channel parameters grows very rapidly and is expensive to track.

In this work, the first issue is resolved by using the suboptimal PM-BPQM decoder because its intermediate channels are all qubit BSCQ channels that are parameterized by two real numbers $(\delta,\gamma)$.
The second issue can also occur with classical channels and is typically resolved by using Monte Carlo DE (known as population dynamics in statistical physics~\cite{mezard2009information}) to approximate the answer efficiently~\cite{daveyldpccommletters}.
The idea of Monte Carlo DE is to approximate distributions over channel parameters by bags of $M$ samples (i.e., a uniform distribution over a length-$M$ list of channel parameters).

Consider a bag $B = \{(\delta_j,\gamma_j)\}_{j=1}^M$ containing $M$ pairs of real numbers corresponding to the parameters of different qubit BSCQ channels.
Then, we define the check-node and bit-node updates of $B$ as follows.
\begin{defn}
   The \emph{check-node update} $B^{\cnop} = \{(\delta_j ',\gamma_j ')\}_{j=1}^M$ of $B$ be constructed as follows.
   For each element $(\delta_{j},\gamma_{j})\in B$, we choose another random element $(\delta_{\pi(j)},\gamma_{\pi(j)})$, where $\pi: [M] \to [M]$ is a uniform random permutation.
   Then, we apply the check-node channel combining operation on the two implied qubit BSCQs.
   The parameters of the resulting qubit BSCQ are given (for $a\in\{0,1\}$) by
   \[ \big(\delta_{a}^{\cnop}(\delta_{j},\gamma_{j},\delta_{\pi (i)},\gamma_{\pi (j)}),\gamma_{a}^{\cnop}(\delta_{j},\gamma_{j},\delta_{\pi (j)},\gamma_{\pi (j)})\big), \]
   with probability $p_{a}^{\cnop}(\delta,\gamma,\delta',\gamma')$.
   The $j$-th value $(\delta_j ', \gamma_j ')$ of $B^{\cnop}$ is set by choosing one of the two according to $p_{a}^{\cnop}$.
\end{defn} 
\begin{defn}
    The \emph{bit-node update} $B^{\vnop} = \{(\delta_j ',\gamma_j ')\}_{j=1}^M$ of $B$ is constructed analogously to the check-node update.
    In particular, the steps are identical but all expressions use the $\vnop$ superscript rather than the $\cnop$ superscript.
\end{defn}

Now, we consider the design of an $(N,K)$ polar code for a qubit BSCQ with parameters $(\delta,\gamma)$.
We implement the DE design of the code using the following steps.
\begin{enumerate}
    \item $B_{0,1} \leftarrow \{(\delta_j,\gamma_j)\}_{j=1}^{M}$ with $(\delta_j,\gamma_j) = (\delta,\gamma)$ for $j\in[M]$ 
    \item For $k$ in $\{1,\ldots,\log_2 N\}$:
    \begin{enumerate}
        \item For $i$ in $\{1,\ldots,2^{k-1}\}$:
        \begin{enumerate}
            \item Compute check-node update:  $B_{k,2i-1} \leftarrow B_{k-1,i}^{\cnop}$
            \item Compute bit-node update: $B_{k,2i} \leftarrow B_{k-1,i}^{\vnop}$
            
        \end{enumerate}
    \end{enumerate}
    \item For $i$ in $\{1,\ldots,N\}$:
    \begin{enumerate}
        \item Using  $B_{n,i} \rightarrow \{(\delta_j,\gamma_j)\}_{j=1}^M$
        \item Compute: $\epsilon_i \leftarrow \frac{1}{M} \sum_{j=1}^{M} \delta_j$
        
    \end{enumerate}
    \item For a length-$N$ polar code with $K$ information bits, let $\mathcal{A} = \{i\in  [N] \mid \epsilon_i \leq \alpha \}$ and choose $\alpha$ so $|\mathcal{A}| = K$.

\end{enumerate}

We note that the DE for PM-BPQM decoding of the effective channel $W_N^{(i)}$ assumes that we make hard decisions about the channel reliability at each stage of decoding (e.g., it measures the second qubit after applying the $C$ or $V$ unitary at each stage).
Under this assumption, $B_{n,i}$ approximates the distribution of the channel parameters seen when decoding $U_i$ given the observation from $W_N^{(i)}$.
Since the Helstrom error rate for a qubit BSCQ with parameters $(\delta,\gamma)$ is $\delta$, the expected Helstrom error rate for $W_N^{(i)}$ under PM-BPQM decoding is approximated by $\epsilon_i$.
Thus, the design method approximates the expected error rate of each effective channel and then chooses the $K$ information bits whose effective channels have the smallest error rates.

\section{Numerical Results for Polar Code Design}

\begin{figure}[t]
\vspace{-2mm}
\centering
\begin{tikzpicture}

\definecolor{darkgray176}{RGB}{176,176,176}
\definecolor{darkorange25512714}{RGB}{255,127,14}
\definecolor{forestgreen4416044}{RGB}{44,160,44}
\definecolor{lightgray204}{RGB}{204,204,204}
\definecolor{steelblue31119180}{RGB}{31,119,180}

\begin{axis}[
width=3.35in,
height=2.5in,
legend cell align={left},
legend style={
  fill opacity=0.8,
  draw opacity=1,
  text opacity=1,
  at={(0.03,0.97)},
  anchor=north west,
  draw=lightgray204,
  font=\small
},
tick align=outside,
tick pos=left,
x grid style={darkgray176},
xlabel={$\gamma$},
xmajorgrids,
xmin=-0.01225, xmax=0.25725,
xminorgrids,
xtick style={color=black},
xticklabel style={
            /pgf/number format/precision=2,
            /pgf/number format/fixed, /pgf/number format/fixed zerofill
        },
y grid style={darkgray176},
ylabel={Code Rate},
ylabel near ticks,
ymajorgrids,
ymin=0.43, ymax=0.60,
yminorgrids,
ytick style={color=black},
yticklabel style={
            /pgf/number format/precision=2,
            /pgf/number format/fixed, /pgf/number format/fixed zerofill
        }
]
\addplot [mark=o,ultra thick, steelblue31119180]
table {%
0 0.4833984375
0.035 0.4775390625
0.105 0.470703125
0.14 0.47265625
0.175 0.482421875
0.21 0.51171875
0.245 0.5927734375
};
\addlegendentry{PM-BPQM:UB}
\addplot [mark=o,ultra thick, darkorange25512714]
table {%
0 0.4541015625
0.035 0.447265625
0.105 0.439453125
0.14 0.4384765625
0.175 0.451171875
0.21 0.4765625
0.245 0.5537109375
};
\addlegendentry{PM-BPQM:NCUB}
\addplot [mark=o,ultra thick, forestgreen4416044]
table {%
0 0.482421875
0.035 0.4853515625
0.105 0.482421875
0.14 0.4833984375
0.175 0.4833984375
0.21 0.4775390625
0.245 0.48046875
};
\addlegendentry{MF:UB}
\end{axis}
\end{tikzpicture}
  \vspace{-7mm}
  \caption{Comparison of PM-BPQM and MF polar decoder for $N=1024$ on qubit BSCQ channels with $\delta=0.07$ and variable $\gamma$ under a union-bound block-error constraint of 0.1.}
 \label{pmbpqm vs hard dec}
\end{figure}
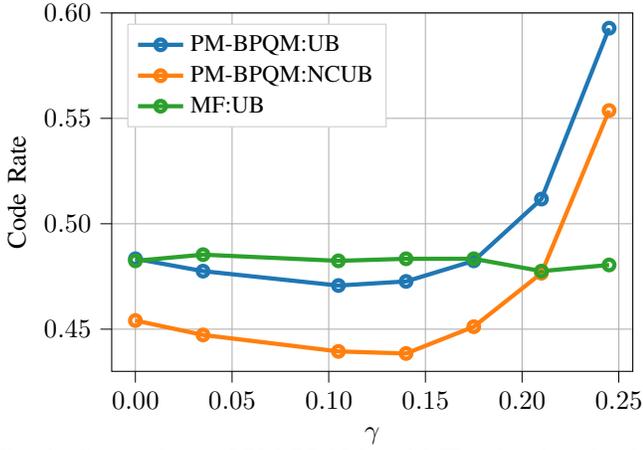

In Fig.~\ref{pmbpqm vs hard dec} and Fig.~\ref{pmbpqm vs hard dec1}, we plot the code rate achievable by the PM-BPQM decoder for length-$1024$ polar codes.
Our results consider channels with $\delta \in \{0.07,0.09\}$ and a range of $\gamma$.
We also compare the results with a measure first (MF) strategy that uses a classical polar code designed for the binary symmetric channel.
Its curve is labeled MF:UB because it uses the classical union bound.
All codes are designed under a union-bound constraint on the block-error probability of 0.1.

Since the union bound for classical and quantum events differs roughly by a factor of 4~\cite{gao2015quantum,khabbazi2019union,o2022quantum}, a fair comparison is challenging and we make two different unfair comparisons.
The curve labeled PM-BPQM:UB ignores the factor of 4 and uses the classical union bound to enforce the block error constraint.
Comparing it with the MF:UB curve is somewhat unfair to MF strategy.
The curve labeled PMBPQM:NCUB uses Gao's bound~\cite{gao2015quantum}.
Comparing it with the MF:UB curve gives MF strategy an unfair advantage.
In both cases, the PM-BPQM decoder achieves a higher rate than the hard-decision decoder for large values of $\gamma$ but the transition point increases for the non-commutative union bound.

\ifarxiv
In Appendix~\ref{app:pol_pmbpqm}, we also plot
\fi
the channel error rates as a function of block length in order to visualize polarization.
The capacity of the qubit BSCQ is compared with the hard-decision capacity in
\ifarxiv
Appendix~\ref{app:qbscq_cap}.
\fi

\section{Simulation of the PM-BPQM Polar Decoder}

The PM-BPQM decoding process is assumed to measure the reliability information of intermediate channels during decoding~\cite{brandsen2022belief}.
While these measurements do not affect the performance of the targeted information bit (e.g., the first non-frozen bit or the root node of an LDPC code tree), they do disturb the quantum state and hurt the performance of later bits (e.g., the second non-frozen bit).
It is well-known that this degradation can be avoided by delaying the intermediate measurements using the quantum principle of deferred measurement~\cite{Renes-njp17,rengaswamy2021belief}.

The cost of deferring measurements is that all unitary operations done after a deferred measurement must be implemented as conditional unitary operations that depend on the system that was not measured.
This results in conditional unitary operations that depend on many qubits.
While such operations are difficult to realize on physical quantum computers, they can be implemented with relatively low complexity in a classical simulation of a quantum computer.
This is the approach we use in our simulation code.
For a physical quantum computer, Renes and Piveteau recently described another approach that achieves quadratic complexity by using reliability registers in the decoder to reduce the burden of conditioning~\cite{piveteau2022quantum}.

\ifarxiv
In Appendix~\ref{pmbpqm-deferred-bscq},
we describe the decoding process with deferred measurements for length-$N$ polar code. We provide descriptions for both BPQM on the pure-state channel and PM-BPQM on a qubit BSCQ channel for length-4 polar code.
We compare the performance of this decoder (by simulating the whole quantum system) with the DE calculation (which only uses the expressions in Lemmas~\ref{lem:bscq-checknode} and~\ref{bscq-bitnode}.
\fi

\begin{figure}[t]
\centering
\begin{tikzpicture}

\definecolor{crimson2143940}{RGB}{214,39,40}
\definecolor{darkgray176}{RGB}{176,176,176}
\definecolor{darkorange25512714}{RGB}{255,127,14}
\definecolor{darkturquoise23190207}{RGB}{23,190,207}
\definecolor{forestgreen4416044}{RGB}{44,160,44}
\definecolor{goldenrod18818934}{RGB}{188,189,34}
\definecolor{gray127}{RGB}{127,127,127}
\definecolor{mediumpurple148103189}{RGB}{148,103,189}
\definecolor{orchid227119194}{RGB}{227,119,194}
\definecolor{sienna1408675}{RGB}{140,86,75}
\definecolor{steelblue31119180}{RGB}{31,119,180}

\begin{axis}[
width=3.35in,
height=2.5in,
tick align=outside,
tick pos=left,
x grid style={darkgray176},
xlabel={\(\displaystyle \gamma\)},
xmajorgrids,
xmin=-0.014, xmax=0.294,
xminorgrids,
xtick style={color=black},
y grid style={darkgray176},
ylabel={Bit-Error Rate},
ymajorgrids,
ymin=-0.0204390514046846, ymax=0.454592335781175,
yminorgrids,
ytick style={color=black}
]
\addplot [semithick, steelblue31119180]
table {%
0 0.41611392
0.05 0.416107087980236
0.1 0.416055352688864
0.15 0.416218133917444
0.2 0.415673730184305
0.25 0.416084367235248
0.28 0.415791598568638
};
\addlegendentry{DE1}
\addplot [semithick, darkorange25512714]
table {%
0 0.29373949110073
0.05 0.292254821702935
0.1 0.281655397785379
0.15 0.264263116496144
0.2 0.244529641218262
0.25 0.221227095623152
0.28 0.207185498064167
};
\addlegendentry{DE2}
\addplot [semithick, forestgreen4416044]
table {%
0 0.295526958968565
0.05 0.292315044722869
0.1 0.277933394734789
0.15 0.257314213675103
0.2 0.233406815969805
0.25 0.20007394631294
0.28 0.173332863713652
};
\addlegendentry{DE3}
\addplot [semithick, crimson2143940]
table {%
0 0.085900968003743
0.05 0.0858218951612495
0.1 0.0836024881141929
0.15 0.0776897803265689
0.2 0.0680684999384584
0.25 0.0505035961107294
0.28 0.0345473306703641
};
\addlegendentry{DE4}
\addplot [semithick, mediumpurple148103189]
table {%
0 0.298131547572495
0.05 0.291309448137611
0.1 0.278476628290467
0.15 0.259485560827571
0.2 0.223789086899156
0.25 0.180548368411879
0.28 0.151023548144254
};
\addlegendentry{DE5}
\addplot [semithick, sienna1408675]
table {%
0 0.07510560955669
0.05 0.0736870739796811
0.1 0.0708038016072927
0.15 0.0669852695381367
0.2 0.0564293270541926
0.25 0.0397891957457863
0.28 0.0264907276012849
};
\addlegendentry{DE6}
\addplot [semithick, orchid227119194]
table {%
0 0.0550302335900342
0.05 0.0547998668337772
0.1 0.0532266899200449
0.15 0.0569867758486398
0.2 0.0478946494518695
0.25 0.0344314620505026
0.28 0.0212650272642687
};
\addlegendentry{DE7}
\addplot [semithick, gray127]
table {%
0 0.00269640978548158
0.05 0.00303683049865966
0.1 0.00286795438410536
0.15 0.00360888642055979
0.2 0.00293450831653563
0.25 0.0027351086029288
0.28 0.00115355834729747
};
\addlegendentry{DE8}
\addplot [semithick, steelblue31119180, only marks, mark=*]
table {%
0 0.433
0.05 0.406013285249116
0.1 0.428595490954372
0.15 0.414336373157761
0.2 0.416441600000004
0.25 0.414699701210866
0.28 0.412961335532298
};
\addlegendentry{Sim1}
\addplot [semithick, darkorange25512714, only marks, mark=*]
table {%
0 0.282
0.05 0.264736732901107
0.1 0.28176616687291
0.15 0.273592669028719
0.2 0.248221455891353
0.25 0.219372439052646
0.28 0.201909959924518
};
\addlegendentry{Sim2}
\addplot [semithick, forestgreen4416044, only marks, mark=*]
table {%
0 0.295
0.05 0.269795763598824
0.1 0.277908498462383
0.15 0.261594723204713
0.2 0.242071067005461
0.25 0.19632916788528
0.28 0.168295409670801
};
\addlegendentry{Sim3}
\addplot [semithick, crimson2143940, only marks, mark=*]
table {%
0 0.0810000000000003
0.05 0.0745038770671372
0.1 0.0786463345149953
0.15 0.0866199642699144
0.2 0.0737508673807332
0.25 0.0497030286177137
0.28 0.0319616638441536
};
\addlegendentry{Sim4}
\addplot [semithick, mediumpurple148103189, only marks, mark=*]
table {%
0 0.297
0.05 0.280229760990181
0.1 0.286066772961706
0.15 0.264559405582452
0.2 0.230800768256733
0.25 0.184167118833631
0.28 0.139582682632025
};
\addlegendentry{Sim5}
\addplot [semithick, sienna1408675, only marks, mark=*]
table {%
0 0.065000000000001
0.05 0.0626498001356725
0.1 0.0659199856496859
0.15 0.0729172521193913
0.2 0.0577866364620215
0.25 0.0397959333048028
0.28 0.0236027186131887
};
\addlegendentry{Sim6}
\addplot [semithick, orchid227119194, only marks, mark=*]
table {%
0 0.0470000000000013
0.05 0.0570053573064121
0.1 0.0457903319942032
0.15 0.0597027218577017
0.2 0.0559874367258089
0.25 0.0344985208195615
0.28 0.0167858542688323
};
\addlegendentry{Sim7}
\addplot [semithick, gray127, only marks, mark=*]
table {%
0 0.00200000000000222
0.05 0.00187231857370962
0.1 0.00272817160121858
0.15 0.00425992280338439
0.2 0.00299559553232608
0.25 0.00334979116955895
0.28 0.00115328437649088
};
\addlegendentry{Sim8}
\legend{}
\end{axis}

\end{tikzpicture}
    \vspace{-3.5mm}
    \caption{Comparison of bit-error rate between DE analysis (solid lines) and simulated decoder (circles) for the effective channels of bits $u_1,\ldots,u_8$ of a length-8 polar code over qubit BSCQ channels with $\delta=0.1$ and variable $\gamma$.}
 \label{devsdecoder_bscq}
 \vspace{1.5mm}
\end{figure}
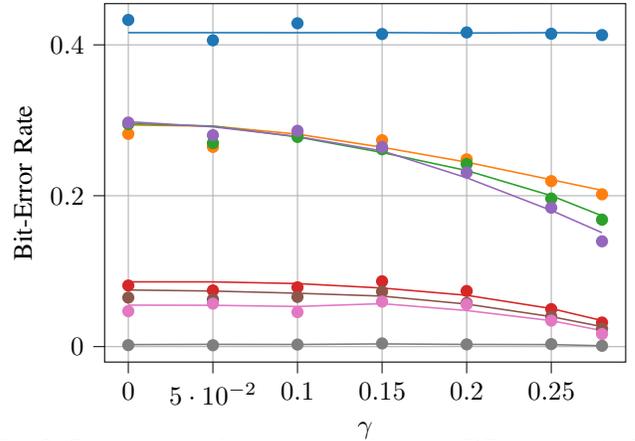

The results can be found in Fig.~\ref{devsdecoder_bscq} where we plot the Helstrom error rate corresponding to each channel between DE output and the PM-BPQM based
\ifarxiv
polar decoder discussed in Appendix~\ref{pmbpqm-deferred-bscq}.
\fi

\begin{figure}[t]
\centering
\begin{tikzpicture}

\definecolor{darkgray176}{RGB}{176,176,176}
\definecolor{darkorange25512714}{RGB}{255,127,14}
\definecolor{forestgreen4416044}{RGB}{44,160,44}
\definecolor{lightgray204}{RGB}{204,204,204}
\definecolor{steelblue31119180}{RGB}{31,119,180}

\begin{axis}[
width=3.35in,
height=2.5in,
legend cell align={left},
legend style={
  fill opacity=0.8,
  draw opacity=1,
  text opacity=1,
  at={(0.03,0.97)},
  anchor=north west,
  draw=lightgray204
},
tick align=outside,
tick pos=left,
x grid style={darkgray176},
xlabel={$\gamma$},
xmajorgrids,
xmin=-0.0135, xmax=0.2835,
xminorgrids,
xtick style={color=black},
xticklabel style={
            /pgf/number format/fixed, 
            /pgf/number format/precision=2,
            /pgf/number format/fixed zerofill
        },
y grid style={darkgray176},
ylabel={Code Rate},
ylabel near ticks,
ymajorgrids,
ymin=0.34, ymax=0.51,
yminorgrids,
ytick style={color=black},
yticklabel style={
            /pgf/number format/fixed, 
            /pgf/number format/precision=2,
            /pgf/number format/fixed zerofill
        }
]
\addplot [mark=o,ultra thick, steelblue31119180]
table {%
0 0.4111328125
0.045 0.412109375
0.135 0.3994140625
0.18 0.4091796875
0.225 0.4296875
0.27 0.501
};
\addlegendentry{PM-BPQM:UB}
\addplot [mark=o,ultra thick, darkorange25512714]
table {%
0 0.380859375
0.045 0.3818359375
0.135 0.369140625
0.18 0.3779296875
0.225 0.396484375
0.27 0.464
};
\addlegendentry{PM-BPQM:NCUB}
\addplot [mark=o, ultra thick, forestgreen4416044]
table {%
0 0.4140625
0.045 0.412109375
0.135 0.412109375
0.18 0.4150390625
0.225 0.412109375
0.27 0.413
};
\addlegendentry{MF:UB}
\end{axis}

\end{tikzpicture}
  \vspace{-2.5mm}
  \caption{Comparison of PM-BPQM and MF polar decoder with $N=1024$ on qubit BSCQ channels with $\delta=0.09$ and variable $\gamma$ under a union-bound block-error constraint of 0.1.}
 \label{pmbpqm vs hard dec1}
\end{figure}
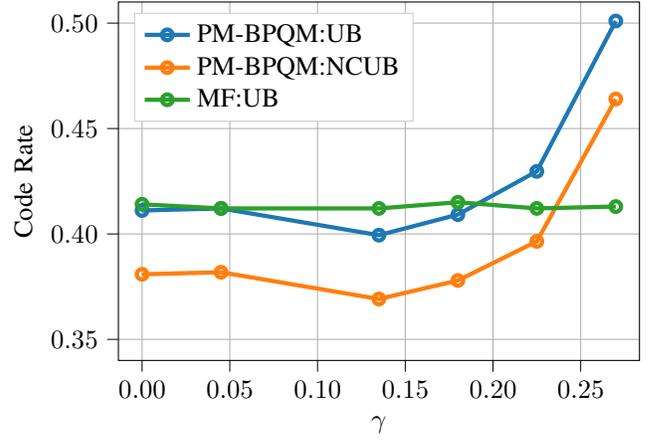



\section{Conclusion}

In this paper, we consider the design and decoding of polar codes on general CQ channels.
Our approach is based on analyzing and implementing the suboptimal PM-BPQM decoder.
On the analysis side, we use DE to design polar codes for general CQ channels under PM-BPQM decoding.
This process can be easily implemented on a classical computer and allows one to explore the achievable trade-off between rate and block-error probability.

We have also implemented the PM-BPQM polar decoder for arbitrary $N$ in Python.
It consists of a classical recursive algorithm that controls a quantum simulator (or quantum computer) to implement the PM-BPQM decoding process.
The code can be found on GitHub in the repository: \vspace{-0.5mm}
\begin{center}\url{https://github.com/Aviemathelec1995/CQ-Polar-BPQM}.
\end{center}
\vspace{0.5mm}

This decoder was used to perform some experiments.
In Fig.~\ref{devsdecoder_bscq}, we compare the simulated decoding performance with the DE prediction for a length-8 polar code and observe a good agreement.
The decoder also allows us to analyze the block error rate without resorting to union bounds.
For the length-8 code, we use the 4-th, 6-th, 7-th and 8-th input bits as information bits.
We estimate the block error rate using 1000 blocks over the BSCQ channel with $(\delta,\gamma)=(0.05,0.15)$.
Its value is roughly 0.07 when we use the frozen set $(u_1,u_2,u_3,u_5)=(1,1,1,1)$ and random information symbols.
For comparison, the error rates of the individual channels $(u_4,u_6,u_7,u_8)$
are computed using DE and they are 0.0178, 0.0146, 0.0123, and 0.0003, respectively.
The classical union bound on block error rate equals the sum of the individual channel error rates (i.e., roughly $0.045$).
Thus, the observed block error rate is roughly twice the classical union bound and less than the factor of $4$ worst-case increase allowed by Gao's bound~\cite{gao2015quantum}.


\IEEEtriggeratref{12}

\clearpage

\printbibliography

\clearpage

\begin{appendices}
\onecolumn

\section{Qubit BSCQ Capacity Comparison}
\label{app:qbscq_cap}

For the qubit BSCQ defined by $W(z)=\sigma_{x}^{z}\rho(\delta,\gamma)\sigma_{x}^{z}$, the channel capacity $C_{\delta,\gamma}$ is given by
\begin{align}
    C(\delta,\gamma) \coloneqq S\left(\frac{\rho(\delta,\gamma)+\sigma_{x}\rho(\delta,\gamma)\sigma_{x}}{2} \right)-S(\rho(\delta,\gamma) ),
\end{align}
where $S(\rho) = - \Tr (\rho \ln \rho)$ is the von Neumann entropy of $\rho$. The eigenvalues of the density matrix $\rho(\delta,\gamma)$ can be obtained as $\lambda, 1-\lambda$ where
\begin{align}
    \lambda=\frac{1+\sqrt{1-4(\delta(1-\delta)-|\gamma|^{2})}}{2}.
\end{align}
Assuming $\gamma$ is real, we can write the capacity $C(\delta,\gamma)$ as 
\begin{align}
    h_{2}\left(\frac{1+2\gamma}{2}\right)- h_{2}\left(\frac{1+\sqrt{1-4(\delta(1-\delta)-|\gamma|^{2})}}{2}\right),
\end{align}
where $h_{2}(\cdot)$ stands for binary entropy function.
For our experiments, we choose qubit BSCQ channels with fixed  $\delta$ fixed and increase $\gamma$ over some range.
For a fixed $\delta$, it is easy to check that $C_{\delta,\gamma}$ is monotonically increasing in $\gamma$.  When $|\gamma|^{2}=\delta(1-\delta)$, the channel becomes a pure-state channel while for $\gamma=0$ the channel becomes a BSC. 

\section{Polarization of PM-BPQM Decoding}
\label{app:pol_pmbpqm}

\begin{figure}[t]
\centering
  \input{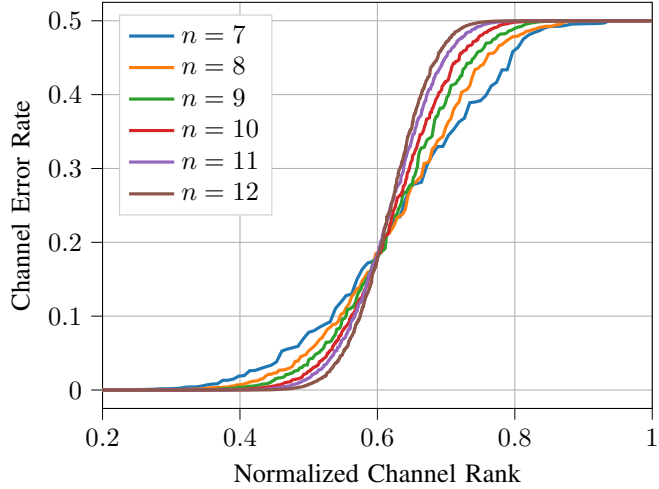}
 \caption{DE polar design curve for PM-BPQM decoding of a length-$2^{n}$ code on the qubit channel with $(\delta,\gamma)=(0.08,0.05)$.}
 \label{polarization}
\end{figure}

 In Fig.~\ref{polarization}, we plot a polar code design curve for length $N=2^{n}$ polar codes.
 This curve shows the average Helstrom error rate for PM-BPQM decoding (obtained from DE) sorted into increasing order.
 For the DE results, we used parameters $M=10000$, $\delta=0.08$, $\gamma=0.05$ and $n\in\{7,8,9,10,11,12\}$.
 This result shows that polarization appears to occur with PM-BPQM even though it is a suboptimal decoding algorithm.
 In particular, the fraction of ``good'' channels is increasing with $n$ and not tending to zero.

\section{Deferred Measurement Polar Decoding}
\label{pmbpqm-deferred-bscq}

In this section, we will analyze the PM-BPQM SC decoder with deferred measurements to decode polar code when the message is transmitted over a qubit BSCQ channel.
In the case of deferred measurements, we form the factor graph associated with each information bit (say $u_i$) and, after applying the check-node or the bit-node unitaries,  we do not measure the reliability but instead we form the conditional unitaries coherently accounting for $(\delta,\gamma)$ parameter updates.
Then, we only measure the decision qubit for $u_i$ in the last stage.
After that, we invert the unitary operations to make the overall operation into a binary projective measurement.
This description and our current implementation use naive deferred measurements and do not yet implement the improvement in~\cite{piveteau2022quantum}.

\subsection{Length-4 PM-BPQM Polar Decoder}
To explain the decoding process, we consider a length-4 polar code over a BSCQ channel.  To denote a vector $(a_1,\dots, a_{N})$, we use notation $a_{1}^{N}$.
 We assume each code bit is transmitted over an identical qubit BSCQ channel i.e.  $W(z)=\sigma_{x}^z\rho(\delta,\gamma)\sigma_{x}^z$. In this case, the input bits $u_{1}^{4}=(u_{1},u_{2},u_{3},u_{4})$ are mapped to codeword bits $x_{1}^{4}=(x_{1},x_{2},x_{3},x_{4})$ using the length-4 polar transform $G_{4}$ such that
 \begin{align*}
 \small
     G_{4}=\begin{bmatrix}
     1&0 &0 &0\\
     1&0 &1 &0\\
     1&1 &0 &0\\
     1&1 &1 &1
     \end{bmatrix}.
 \end{align*}
 Let $A_{i}$ be the quantum system associated with qubit $i$, then for the codeword sequence $\underline{x}$, the output is given by 
\begin{align*}
    &W^{A_{1}A_{2}A_{3}A_{4}}  (u_{1}^{4}) \\ &\quad =W^{A_{1}}(x_{1})\otimes W^{A_{2}}(x_{2})\otimes W^{A_{3}}(x_{3})\otimes W^{A_{4}}(x_{4}).
\end{align*}

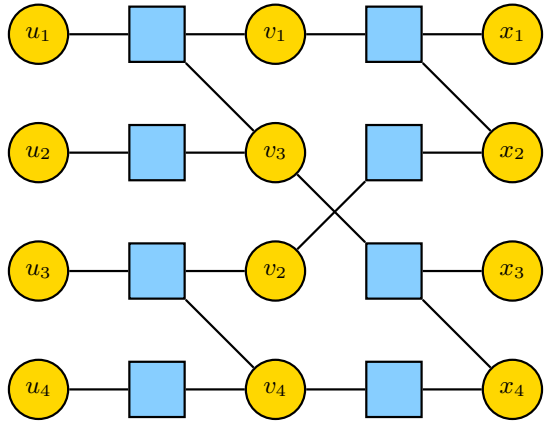
\begin{figure}[b]
    \centering
    \scalebox{1.05}{\begin{tikzpicture}%
    [scale=0.75,var/.style={font=\small,fill=bitcolor,draw,circle,thick,minimum size=7.5mm},%
    factor/.style={font=\small,fill=checkcolor,draw,rectangle,thick,minimum size=7mm},%
    weight/.style={font=\small}]
    
    \node (u1) [var] at (-0.5,0) {$u_1$};
    \node (u2) [var] at (-0.5,-2) {$u_2$};
    \node (u3) [var] at (-0.5,-4) {$u_3$};
    \node (u4) [var] at (-0.5,-6) {$u_4$};
    \node (f1) [factor] at (1.5,0) {};
    \node (f2) [factor] at (1.5,-2) {};
    \node (f3) [factor] at (1.5,-4) {};
    \node (f4) [factor] at (1.5,-6) {};
    \node (v1) [var] at (3.5,0) {$v_1$};
    \node (v2) [var] at (3.5,-2) {$v_3$};
    \node (v3) [var] at (3.5,-4) {$v_2$};
    \node (v4) [var] at (3.5,-6) {$v_4$};
    \node (g1) [factor] at (5.5,0) {};
    \node (g2) [factor] at (5.5,-2) {};
    \node (g3) [factor] at (5.5,-4) {};
    \node (g4) [factor] at (5.5,-6) {};
    \node (x1) [var] at (7.5,0) {$x_1$};
    \node (x2) [var] at (7.5,-2) {$x_2$};
    \node (x3) [var] at (7.5,-4) {$x_3$};
    \node (x4) [var] at (7.5,-6) {$x_4$};
    
    \draw[thick] (x1) -- (g1);      
    \draw[thick] (x2) -- (g1);      
    \draw[thick] (x2) -- (g2);      
    \draw[thick] (x3) -- (g3);      
    \draw[thick] (x4) -- (g3);      
    \draw[thick] (x4) -- (g4);
    \draw[thick] (g1) -- (v1);
    \draw[thick] (g2) -- (v3);
    \draw[thick] (g3) -- (v2);
    \draw[thick] (g4) -- (v4);
    \draw[thick] (v1) -- (f1);      
    \draw[thick] (v2) -- (f1);      
    \draw[thick] (v2) -- (f2);      
    \draw[thick] (v3) -- (f3);      
    \draw[thick] (v4) -- (f3);      
    \draw[thick] (v4) -- (f4);
    \draw[thick] (f1) -- (u1);
    \draw[thick] (f2) -- (u2);
    \draw[thick] (f3) -- (u3);
    \draw[thick] (f4) -- (u4);

    \end{tikzpicture}}
    \vspace{2mm}
    \caption{Decoding factor graph for a length-4 polar code}
    \label{polardec4}
\end{figure}

For simplicity, we discuss the decoder analysis for the case where all the input bits are 0.
Since we have a linear code on a symmetric channel, the result for any other codeword is related by a tensor product of $\sigma_{x}$ operators.
This implies 
\begin{align}
    &W^{A_{1}A_{2}A_{3}A_{4}}(u_{1}^{4})\\ &\quad =\rho^{A_{1}}(\delta,\gamma)\otimes \rho^{A_{2}}(\delta,\gamma)\otimes \rho^{A_{3}}(\delta,\gamma)\otimes \rho^{A_{4}}(\delta,\gamma).
\end{align}
To realize the successive cancellation decoder for decoding bit $u_{i}$, we construct a factor graph with root node $u_{i}$ and decoder output for past bits as $\hat{u}_{1}^{i-1}$ from the decoding factor graph of length-4 polar code depicted in Fig.~\ref{polardec4}. Before describing the decoding of each bit $u_{i}$, we define some additional operators that are necessary to analyze deferred measurement decoding. 

   
          
Let us define the (generalized) swap operator which permutes $N$ qubits according to a permutation $\pi:[N] \to [N]$.
It is uniquely defined by its action on tensor product states, where $\ket{\psi_{i}}$ is the state of the quantum system $A_{i}$, as follows 
\vspace{-0.05mm}
\begin{align}
Sw_{\pi}^{(N)}\ket{\psi_{1}}_{A_{1}}\!\!\otimes\dots\otimes\ket{\psi_{N}}_{A_{N}}\!=\ket{\psi_{\pi(1)}}_{A_{1}}\!\!\otimes\dots\otimes\ket{\psi_{\pi(N)}}_{A_{N}}\!.\!\!\!\!
\end{align}

\vspace{-1mm}

Let $V_{i,j}(\delta_{i},\gamma_{i},\delta_{j},\gamma_{j})$ denote the unitary corresponding to bit-node channel combining for the  $i$-th and $j$-th channels with output states $\rho^{A_{1}}(\delta_{i},\gamma_{i})$ and $\rho^{A_{2}}(\delta_{j},\gamma_{j})$ respectively.
We denote $C_{i,j}^{(N)}$ as the unitary associated with check-node channel combining for the channels $W_{i}^{A_{i}}$ and $W_{j}^{A_{j}}$ since the check node unitary  acting on two qubits does not depend on the parameters of qubit states.
In the joint quantum system with more than two qubits, we assume that the unitaries $V_{i,j}^{(N)}(\delta_{i},\gamma_{i},\delta_{j},\gamma_{j})$ and $C^{(N)}_{i,j}$ act only on qubits $i$ and $j$ and are tensored with identities on the other systems.
Using the swap operator $Sw^{(N)}_{\pi}$ and the unitaries $C$ and $V(\delta_{i},\gamma_{i},\delta_{j},\gamma_{j})$, we can define $C_{i,j}^{(N)}$ and $V_{i,j}^{(N)}(\delta_{i},\gamma_{i},\delta_{j},\gamma_{j})$ by 
\vspace{-0.5mm}
\begin{align}
    V_{i,j}^{(N)}(\delta_{i},\gamma_{i},\delta_{j},\gamma_{j}) &  =Sw^{(N)\dagger}_{\pi_{i,j}}(V(\delta_{i},\gamma_{i},\delta_{j},\gamma_{j})\otimes \mathbb{I}_{2^{N-2}})Sw^{(N)}_{\pi_{i,j}}\!\!\! \\
    C^{(N)}_{i,j} &  =Sw^{(N)\dagger}_{\pi_{i,j}}(C\otimes \mathbb{I}_{2^{N-2}})Sw^{(N)}_{\pi_{i,j}}
\end{align}

\vspace{-1mm}
\noindent
where, for distinct $i,j\in [N]$, $\pi_{i,j}$ is a fixed permutation satisfying $\pi_{i,j} (1) = i$ and $\pi_{i,j}(2)=j$.
 For the joint quantum system with 4 qubits suppose, conditioned on reliability qubits, systems $i,j$ have different $(\delta,\gamma)$ parameters.
 Since the check-node unitary does not depend on the channel parameters, we can still use $C_{i,j}^{(4)}$ to implement the check-node operation.
 However, since the bit-node unitary depends on the channel parameters, we will also need to design conditional a bit-node unitary that conditions on the reliability qubits associated with deferred measurements.
 Assuming the reliability information for systems $A_{3}$ 
 and $A_{4}$ with parameter sets $\{\delta'_{j},\gamma'_{j}\}_{j=0}^{    1}$ and$\{\delta''_{j},\gamma''_{j}\}_{j=0}^{1}$ is held in systems $A_{1}$ and $A_{2}$ respectively, the conditional bit node unitary denoted as $V_{3,4|1,2}^{cond}\Big{(}\{\delta'_{j},\gamma'_{j}\}_{j=0}^{    1},\{\delta''_{j},\gamma''_{j}\}_{j=0}^{1}\Big{)}$ has the following structure
\begin{align}
    V_{3,4|1,2}^{cond}\Big{(}\{\delta'_{j},\gamma'_{j}\}_{j=0}^{    1},\{\delta''_{j},\gamma''_{j}\}_{j=0}^{1}\Big{)} 
 =  \begin{bmatrix}
   V_{1}& &&\\
   & V_{2}& & \\
   & & V_{3}& \\
   & & & V_{4}
  \end{bmatrix}.
\end{align}
where $V_{1}=V(\delta'_{0},\gamma'_{0},\delta''_{0},\gamma''_{0})$, $V_{2}=V(\delta'_{0},\gamma'_{0},\delta''_{1},\gamma''_{1})$,$V_{3}=V(\delta'_{1},\gamma'_{1},\delta''_{0},\gamma''_{0})$ and $V_{4}=V(\delta'_{1},\gamma'_{1},\delta''_{1},\gamma''_{1})$. We describe the conditional unitary for the system $A_{3}$ and $A_{4}$ only because it has a nice block diagonal structure. It is worth noting that we can still construct conditional unitary for the systems $A_{i}$ and $A_{j}$ with reliability in systems $A_{k}$ and $A_{l}$ for $i,j,k,l\in [4]$,  by using $V^{cond}_{3,4|1,2}$ and the swap operator $Sw_{\pi}^{(4)}$ with an appropriate permutation $\pi$. 

\begin{figure}[t]
    \centering
    \scalebox{1.05}{\begin{tikzpicture}%
    [scale=0.75,var/.style={font=\small,fill=bitcolor,draw,circle,thick,minimum size=7.5mm},%
    factor/.style={font=\small,fill=checkcolor,draw,rectangle,thick,minimum size=7mm},%
    weight/.style={font=\small}]
    
    \node (u1) [var] at (-0.5,0) {$u_1$};
    \node (f1) [factor] at (1.5,0) {};
    \node (v1) [var] at (3.5,0) {$v_1$};
    \node (v2) [var] at (3.5,-2) {$v_3$};
    \node (g1) [factor] at (5.5,0) {};
    \node (g3) [factor] at (5.5,-4) {};
    \node (x1) [var] at (7.5,0) {$x_1$};
    \node (x2) [var] at (7.5,-2) {$x_2$};
    \node (x3) [var] at (7.5,-4) {$x_3$};
    \node (x4) [var] at (7.5,-6) {$x_4$};
    
    \draw[->,thick] (x1) -- (g1);      
    \draw[->,thick] (x2) -- (g1);      
    \draw[->,thick] (x3) -- (g3);      
    \draw[->,thick] (x4) -- (g3);      
    \draw[->,thick] (g1) -- (v1);
    \draw[->,thick] (g3) -- (v2);
    \draw[->,thick] (v1) -- (f1);      
    \draw[->,thick] (v2) -- (f1);      
    \draw[->,thick] (f1) -- (u1);

\end{tikzpicture}\hspace{15mm}\begin{tikzpicture}%
    [scale=0.75,var/.style={font=\small,fill=bitcolor,draw,circle,thick,minimum size=7.5mm},%
    factor/.style={font=\small,fill=checkcolor,draw,rectangle,thick,minimum size=7mm},%
    weight/.style={font=\small}]
    
    \node (u1) [var] at (-0.5,0) {$
\hat{u}_1$};
    \node (u2) [var] at (-0.5,-2) {$u_2$};
    \node (f1) [factor] at (1.5,0) {};
    \node (f2) [factor] at (1.5,-2) {};
    \node (v1) [var] at (3.5,0) {$v_1$};
    \node (v2) [var] at (3.5,-2) {$v_3$};
    \node (g1) [factor] at (5.5,0) {};
    \node (g3) [factor] at (5.5,-4) {};
    \node (x1) [var] at (7.5,0) {$x_1$};
    \node (x2) [var] at (7.5,-2) {$x_2$};
    \node (x3) [var] at (7.5,-4) {$x_3$};
    \node (x4) [var] at (7.5,-6) {$x_4$};
    
    \draw[->,thick] (x1) -- (g1);      
    \draw[->,thick] (x2) -- (g1);      
    \draw[->,thick] (x3) -- (g3);      
    \draw[->,thick] (x4) -- (g3);      
    \draw[->,thick] (g1) -- (v1);
    \draw[->,thick] (g3) -- (v2);
    \draw[->,thick] (v1) -- (f1);      
    \draw[->,thick] (v2) -- (f2);      
    \draw[<-,thick] (f1) -- (u1);
    \draw[->,thick] (f1) -- (u2);
    \draw[->,thick] (f2) -- (u2);

\end{tikzpicture}}
    \vspace{2mm}
    \caption{Decoding factor graph for $u_1$ (left) and $u_2$ (right) of a length-4 polar code}
    \label{polar4dec1}
\end{figure}
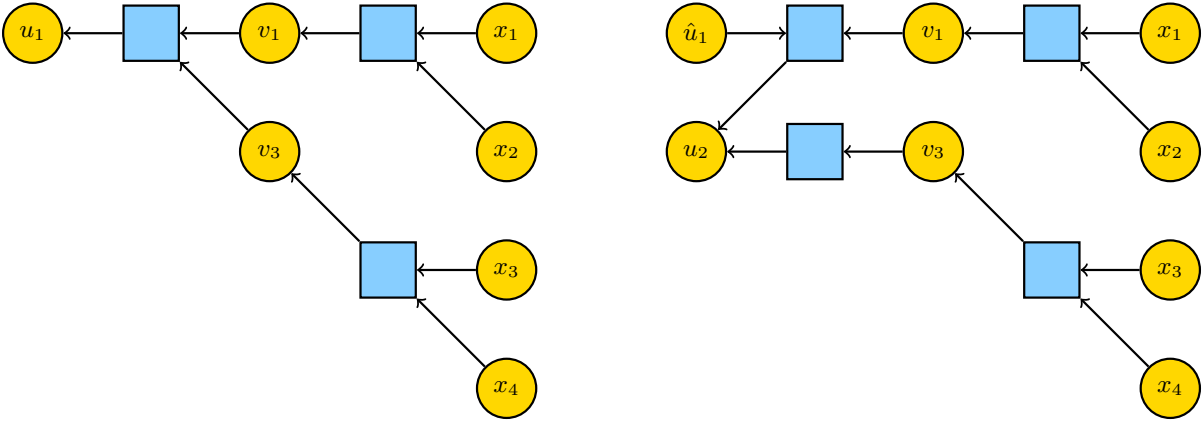


\subsubsection{Decoding bit $u_{1}$}

If $u_{1}$ is an information bit, then we decode it based on the decoding factor graph for bit $u_{1}$ (see Fig.~\ref{polar4dec1}).
Then, we apply the check-node unitaries $C_{1,2}^{(4)}$ and $C_{3,4}^{(4)}$ on $W^{A_{1}A_{2}A_{3}A_{4}}(u_{1}^{4})$ to produce the new state $W_{1}^{A_{1}A_{2}A_{3}A_{4}}(u_{1}^{4})$ given by
\begin{align*}
   W_{1}^{A_{1}A_{2}A_{3}A_{4}}(u_{1}^{4}) &\ =C_{3,4}^{(4)} C_{1,2}^{(4)} W^{A_{1}A_{2}A_{3}A_{4}}(u_{1}^{4}) (C_{1,2}^{(4)})^{\dagger} (C_{3,4}^{(4)})^{\dagger}\\
   &\ \rightarrow (\sum_{j}p_{j}\rho^{A_{1}}(\delta_{j}^{\cnop},\gamma_{j}^{\cnop})\otimes \ketbra{j}{j}_{A_{2}}) \\
   &\ \quad \otimes (\sum_{j}p_{j}\rho^{A_{3}}(\delta_{j}^{\cnop},\gamma_{j}^{\cnop})\otimes \ketbra{j}{j}_{A_{4}})
\end{align*}
Even though we now have a stochastic mixture of qubit states in systems $A_{1}$ and $A_{3}$, determined by the reliability information in systems $A_{2}$ and $A_{4}$ respectively, we just apply the unitary  $C_{1,3}^{(4)}$ on $W_{1}^{A_{1}A_{2}A_{3}A_{4}}(u_{1}^{4})$ because the check node unitary $C$ does not depend on the channel parameters.
This produces the new state $W_{2}^{A_{1}A_{2}A_{3}A_{4}}(u_{1}^{4})$.

If $u_1$ is an information bit, then we would now measure qubit 1 in $W_{2}^{A_{1}A_{2}A_{3}A_{4}}(u_{1}^{4})$ to make the decision.
Regardless, we can compute the error rate of the effective channel for qubit 1 by analyzing this measurement.
Since the symmetry operator for this channel is $\sigma_{x}$, the Helstrom measurement corresponds to projection in the $\sigma_{z}$ basis.
More formally, the Helstrom measurement becomes $\hat{\Pi}_1 =\{\Pi_{1},\mathbb{I}_{16}-\Pi_{1}\}$ where $\Pi_{1}=\ketbra{0}{0}\otimes \mathbb{I}_{8}$.
Thus, the error rate for the effective channel for $u_1$ is 
\begin{align*}
    p_{u_{1}}=1-\text{Tr}(\Pi_{1}W_{2}^{A_{1}A_{2}A_{3}A_{4}}(u_{1}^{4}).
\end{align*}
Depending on the measurement outcome, if hard decision estimate $\hat{u}_1=1$, then we apply $\sigma_{x}$ on the first qubit of $W_{2}^{A_{1}A_{2}A_{3}A_{4}}(u_{1}^{4})$.

\subsubsection{Decoding bit $u_{2}$}
If $u_2$ is an information bit, then we decode it based on the decoding factor graph for bit $u_2$ depicted in Fig.~\ref{polar4dec1}. Like polar decoding for classical BMS channel \cite{arikan2009channel}, we assume the estimate  $\hat{u}_{1}$ is correct. First we apply inverse of the unitary $C_{1,3}^{(4)}$ on $W_{2}^{A_{1}A_{2}A_{3}A_{4}}(u_1^4)$ to produce new state $W_{3}^{A_{1}A_{2}A_{3}A_{4}}(u_1^4)$ as
\begin{align*}
     W_{3}^{A_{1}A_{2}A_{3}A_{4}}(u_1^4)  \quad =(C_{1,3}^{(4)})^{\dagger}W_{2}^{A_{1}A_{2}A_{3}A_{4}}(u_1^4)C_{1,3}^{(4)}.
\end{align*}
Next, we apply $Sw^{(4)}_{\pi'}$ on $W_{3}^{A_{1}A_{2}A_{3}A_{4}}(u_1^4)$  such that $\pi'(1)=3,\pi'(2)=1,\pi'(3)=4,\pi'(4)=2$ and then apply conditional bit node unitary $V_{3,4|1,2}^{cond}\Big{(}\{\delta_{j}^{\cnop},\gamma_{j}^{\cnop}\}_{j=0}^{1},\{\delta_{j}^{\cnop},\gamma_{j}^{\cnop}\}_{j=0}^{1}\Big{)}$ to produce new state $W_{4}^{A_{1}A_{2}A_{3}A_{4}}(\underline{u})$.
In $W_{4}^{A_{1}A_{2}A_{3}A_{4}}(u_1^4)$, the entire information is compressed in system $A_{3}$ so we apply Helstrom measurement using $\hat{\Pi}_{3}=\{\Pi_{3},\mathbb{I}_{16}-\Pi_{3}\}$ where $\Pi_{3}=\mathbb{I}_{4}\otimes \ketbra{0}{0}\otimes\mathbb{I}_{2}$.
Hence, the error for estimating the bit $u_{2}$ becomes 
\begin{align}
    p_{u_{2}}=1-\text{Tr}(\Pi_{3}W_{4}^{A_{1}A_{2}A_{3}A_{4}}(u_1^4))
\end{align}
Finally, depending on the measurement outcome,  we apply $\sigma_{x}$ on $W_{4}^{A_{1}A_{2}A_{3}A_{4}}(u_{1}^{4})$.
\begin{figure}[t]
    \centering
    \scalebox{1.05}{\begin{tikzpicture}%
    [scale=0.75,var/.style={font=\small,fill=bitcolor,draw,circle,thick,minimum size=7.5mm},%
    factor/.style={font=\small,fill=checkcolor,draw,rectangle,thick,minimum size=7mm},%
    weight/.style={font=\small}]
    
    \node (u1) [var] at (-0.5,0) {$
\hat{u}_1$};
    \node (u2) [var] at (-0.5,-2) {$\hat{u}_2$};
    \node (u3) [var] at (-0.5,-4) {$u_3$};
    \node (f1) [factor] at (1.5,0) {};
    \node (f2) [factor] at (1.5,-2) {};
    \node (f3) [factor] at (1.5,-4) {};
    \node (v1) [var] at (3.5,0) {$v_1$};
    \node (v2) [var] at (3.5,-2) {$v_3$};
    \node (v3) [var] at (3.5,-4) {$v_2$};
    \node (v4) [var] at (3.5,-6) {$v_4$};
    \node (g1) [factor] at (5.5,0) {};
    \node (g2) [factor] at (5.5,-2) {};
    \node (g3) [factor] at (5.5,-4) {};
    \node (g4) [factor] at (5.5,-6) {};
    \node (x1) [var] at (7.5,0) {$x_1$};
    \node (x2) [var] at (7.5,-2) {$x_2$};
    \node (x3) [var] at (7.5,-4) {$x_3$};
    \node (x4) [var] at (7.5,-6) {$x_4$};
    
    \draw[->,thick] (x1) -- (g1);      
    \draw[<-,thick] (x2) -- (g1); 
    \draw[->,thick](u2) -- (f1);
    \draw[->,thick] (x2) -- (g2);      
    \draw[->,thick] (x3) -- (g3);      
    \draw[<-,thick] (x4) -- (g3);      
    \draw[->,thick] (x4) -- (g4);
    \draw[<-,thick] (g1) -- (v1);
    \draw[->,thick] (g2) -- (v3);
    \draw[<-,thick] (g3) -- (v2);
    \draw[->,thick] (g4) -- (v4);
    \draw[<-,thick] (v1) -- (f1);      
    \draw[<-,thick] (v2) -- (f2);      
    \draw[->,thick] (v3) -- (f3);      
    \draw[->,thick] (v4) -- (f3);      
    \draw[<-,thick] (f1) -- (u1);
    \draw[<-,thick] (f2) -- (u2);
    \draw[->,thick] (f3) -- (u3);

\end{tikzpicture}\hspace{15mm}\begin{tikzpicture}%
    [scale=0.75,var/.style={font=\small,fill=bitcolor,draw,circle,thick,minimum size=7.5mm},%
    factor/.style={font=\small,fill=checkcolor,draw,rectangle,thick,minimum size=7mm},%
    weight/.style={font=\small}]
    
    \node (u1) [var] at (-0.5,0) {$
\hat{u}_1$};
    \node (u2) [var] at (-0.5,-2) {$\hat{u}_2$};
    \node (u3) [var] at (-0.5,-4) {$\hat{u}_3$};
    \node (u4) [var] at (-0.5,-6) {$u_4$};
    \node (f1) [factor] at (1.5,0) {};
    \node (f2) [factor] at (1.5,-2) {};
    \node (f3) [factor] at (1.5,-4) {};
    \node (f4) [factor] at (1.5,-6) {};
    \node (v1) [var] at (3.5,0) {$v_1$};
    \node (v2) [var] at (3.5,-2) {$v_3$};
    \node (v3) [var] at (3.5,-4) {$v_2$};
    \node (v4) [var] at (3.5,-6) {$v_4$};
    \node (g1) [factor] at (5.5,0) {};
    \node (g2) [factor] at (5.5,-2) {};
    \node (g3) [factor] at (5.5,-4) {};
    \node (g4) [factor] at (5.5,-6) {};
    \node (x1) [var] at (7.5,0) {$x_1$};
    \node (x2) [var] at (7.5,-2) {$x_2$};
    \node (x3) [var] at (7.5,-4) {$x_3$};
    \node (x4) [var] at (7.5,-6) {$x_4$};
    
    \draw[->,thick] (x1) -- (g1);      
    \draw[<-,thick] (x2) -- (g1);  
     \draw[->,thick](u2) -- (f1);
    \draw[->,thick] (x2) -- (g2);      
    \draw[->,thick] (x3) -- (g3);      
    \draw[<-,thick] (x4) -- (g3);      
    \draw[->,thick] (x4) -- (g4);
    \draw[<-,thick] (g1) -- (v1);
    \draw[->,thick] (g2) -- (v3);
    \draw[<-,thick] (g3) -- (v2);
    \draw[->,thick] (g4) -- (v4);
    \draw[<-,thick] (v1) -- (f1);      
    \draw[<-,thick] (v2) -- (f2);      
    \draw[->,thick] (v3) -- (f3);      
    \draw[<-,thick] (u4) -- (f3);      
    \draw[->,thick] (v4) -- (f4);
    \draw[<-,thick] (f1) -- (u1);
    \draw[<-,thick] (f2) -- (u2);
    \draw[<-,thick] (f3) -- (u3);
    \draw[->,thick] (f4) -- (u4);

\end{tikzpicture}}
    \vspace{2mm}
    \caption{Decoding factor graph for $u_3$ (left) and $u_4$ (right) of a length-4 polar code}
    \label{polar4dec3}
\end{figure}
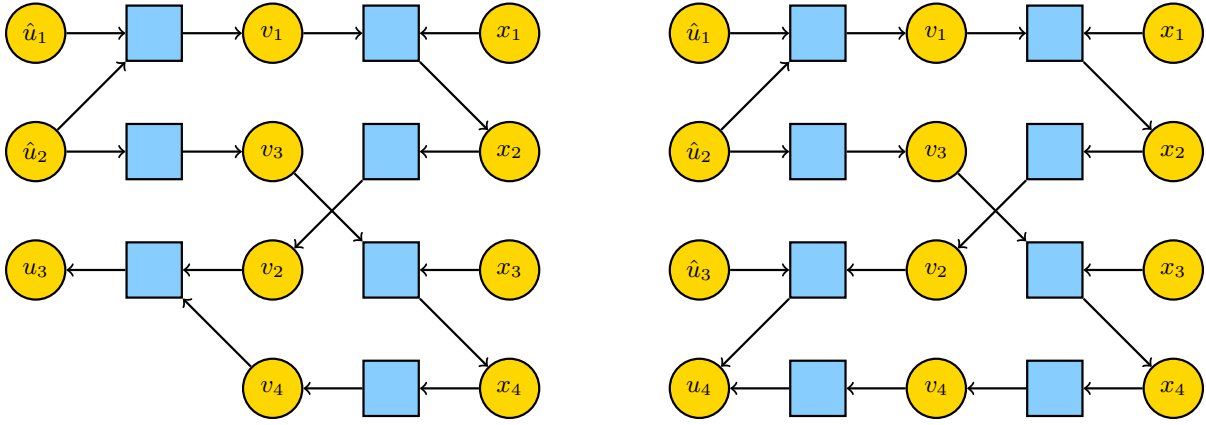


\subsubsection{Decoding bit $u_{3}$}
If $u_3$ is an information bit, then we decode it based on the decoding factor graph for bit $u_3$ depicted in Fig.~\ref{polar4dec3}.
First, we apply $\Big{(}V_{3,4}^{cond}\big{(}\{\delta_{j}^{\cnop},\gamma_{j}^{\cnop}\}_{j=0}^{1},\{\delta_{j}^{\cnop},\gamma_{j}^{\cnop}\}_{j=0}^{1}\big{)}\Big{)}^{\dagger}$ on $W_{4}^{A_{1}A_{2}A_{3}A_{4}}(u_1^4)$ and then we apply $(Sw^{(4)}_{\pi'})^{\dagger}$ to produce the state $W_{5}^{A_{1}A_{2}A_{3}A_{4}}(u_1^4)$. To invert the operation corresponding to unitary the input, we apply  unitary $(C_{3,4}^{(4)})^{\dagger}$ and $(C_{1,2}^{(4)})^{\dagger}$ on $W_{5}^{A_{1}A_{2}A_{3}A_{4}}(u_1^4)$ to produce the state $W_{6}^{A_{1}A_{2}A_{3}A_{4}}(u_1^4)$.  Then we apply bit node unitary $V_{1,2}^{(4)}(\delta,\gamma,\delta,\gamma)$  and $V_{3,4}^{(4)}(\delta,\gamma,\delta,\gamma)$ to produce state $W_{7}^{A_{1}A_{2}A_{3}A_{4}}(u_1^4)$. Finally we apply the check node unitary $C_{1,3}^{(4)}$ on the state  $W_{7}^{A_{1}A_{2}A_{3}A_{4}}(u_1^4)$ to get the new state $W_{8}^{A_{1}A_{2}A_{3}A_{4}}(u_1^4)$. In this case, the information is compressed in $A_{1}$ so we form the Helstrom measurement using $\hat{\Pi}_{1}$. Thus the error rate  for estimating the bit $u_{3}$ is computed as
\begin{align}
    p_{u_{3}}=1-\text{Tr}(\Pi_{1}W_{8}^{A_{1}A_{2}A_{3}A_{4}}(u_1^4))
\end{align}

\subsubsection{Decoding bit $u_{4}$}

If $u_4$ is an information bit, then we decode it based on the decoding factor graph for bit $u_4$ depicted in Fig.~\ref{polar4dec3}. In this case, we apply  $(\sigma^{ \hat{u}_{3}}_{x}\otimes \mathbb{I}_{8})$ and $(C_{1,3}^{(4)})^{\dagger}$ on $W_{8}^{A_{1}A_{2}A_{3}A_{4}}(u_1^4)$. Next we apply $Sw_{\pi'}^{(4)}$ and finally the conditional bit node unitary $V_{3,4|1,2}^{cond}\Big{(}\{\delta_{j}^{\vnop},\gamma_{j}^{\vnop}\}_{j=0}^{1},\{\delta_{j}^{\vnop},\gamma_{j}^{\vnop}\}_{j=0}^{1}\Big{)}$ on the resulting state to produce the final state $W_{9}^{A_{1}A_{2}A_{3}A_{4}}(u_1^4)$. In $W_{9}^{A_{1}A_{2}A_{3}A_{4}}(u_1^4)$ the information is compressed in system $A_{3}$, so apply the Helstrom measurement $\hat{\Pi}_{3}$. The error for estimating the bit $u_{4}$ can be calculated as 
\begin{align}
    p_{u_{4}}=1-\text{Tr}(\Pi_{3}W_{9}^{A_{1}A_{2}A_{3}A_{4}}(u_1^4)).
\end{align}


\subsection{BPQM Decoding for Length-4 code on the Pure-State Channel}

In this section, we will discuss how to design a length-4 polar decoder when the channel is a PSC which is parameterized by angle parameter $\theta$.
We will follow the bit-node and check-node unitary definitions proposed by Renes \cite{Renes-njp17}.
It is worth noting that we can also design bit-node and check-node unitaries by suitably translating the parameter in the PSC domain ($\theta$) to the parameters in the qubit-BSCQ $\{(\delta,\gamma)\}$ domain. 
Consider the PSC where the channel $W$ outputs state $\ket{\pm\theta}$ with $\ket{\pm\theta}=\cos{\frac{\theta}{2}}\ket{0}\pm \sin{\frac{\theta}{2}}\ket{1}$. The bit-node channel combining $W\vnop W'$ outputs $\ket{\theta}\otimes\ket{\theta'}$ or $\ket{-\theta}\otimes\ket{-\theta'}$. There exists a unitary $V(\theta,\theta')$ (\cite[Sec.~3]{Renes-njp17}) with the form 
 \begin{align}
     V(\theta,\theta')=\begin{bmatrix}
     a_{+} & 0 & 0& a_{-}\\
     a_{-} & 0 & 0& -a_{+}\\
    0 & b_{+} & b_{-}& 0\\
     0 & b_{-} & -b_{+}& 0
     \end{bmatrix}
 \end{align}
such that
\begin{align}
    &\ V(\theta,\theta')[W\vnop W'](z)V^{\dagger}(\theta,\theta')\nonumber\\
&\    =\ketbra{(-1)^{z}\theta^{\vnop}}{(-1)^{z}\theta^{\vnop}}\otimes \ketbra{z}{z}.
\end{align}
where $z\in\{0,1\}$, $\theta^{\vnop}=\cos^{-1}{(\cos{\theta}\cos{\theta'})}$, and\\
\begin{align*}
a_{\pm}&=\frac{\frac{1}{\sqrt{2}}\big{(}\cos{\frac{\theta-\theta'}{2}}\pm\cos{\frac{\theta+\theta'}{2}}\big{)}}{\sqrt{1+\cos{\theta}\cos{\theta'}}}, \\\
b_{\pm} &=\frac{\frac{1}{\sqrt{2}}\big{(}\sin{\frac{\theta+\theta'}{2}}\mp\sin{\frac{\theta-\theta'}{2}}\big{)}}{\sqrt{1-\cos{\theta}\cos{\theta'}}}.
\end{align*}

Similarly, for check-node channel combining  $W\cnop W'$, the unitary $U=\text{CNOT}_{1\rightarrow 2}$ acts on $[W\cnop W'](z)$ with $z\in\{0,1\}$ such that
\begin{align*}
    U[W\cnop W'](z)U^{\dagger} \!=\!\! \sum_{j\in\{0,1\}}p_{j}\ketbra{(-1)^{z}\theta^{\cnop}_{j}}{(-1)^{z}\theta^{\cnop}_{j}}\otimes \ketbra{j}{j}.
\end{align*}
Here $\theta_{0}^{\cnop}=\cos^{-1}\big{(}{\frac{\cos{\theta}+\cos{\theta'}}{1+\cos{\theta}\cos{\theta'}}}\big{)}$, $\theta_{1}^{\cnop}=\cos^{-1}\big{(}{\frac{\cos{\theta}-\cos{\theta'}}{1-\cos{\theta}\cos{\theta'}}}\big{)}$, and $p_{0}=\frac{1}{2}(1+\cos{\theta}\cos{\theta'})$, $p_{1}=1-p_{0}$.
The unitaries $V$ and $U$ compress the information to the first qubit. 
We can use these unitaries in the Fig.~\ref{polardec4} polar decoding factor graph to decode the input bits.

In the pure-state case, like the qubit BSCQ scenario,  we write $V_{i,j}^{(N)}(\alpha,\beta)$ for the unitary corresponding to bit-node channel combining for the channels $W_{i}$ and $W_{j}$ with output states $\ket{\alpha}_{A_i}$ and $\ket{\beta}_{A_{j}}$.
We also write $U_{i,j}^{(N)}$ for the unitary associated with the check-node channel combining of the channels $W_{i}$ and $W_{j}$.
We can use $U$ and $V(\alpha,\beta)$ and the swap operator to define $U_{i,j}^{(N)}$ and $V_{i,j}^{(N)}(\alpha,\beta)$ as follows-
\begin{align*}
    V_{i,j}^{(N)}(\alpha,\beta) &\ =Sw^{(N)\dagger}_{\pi_{i,j}}(V(\alpha,\beta)\otimes \mathbb{I}_{2^{N-2}})Sw^{(N)}_{\pi_{i,j}}\\
    U^{(N)}_{i,j} &\ =Sw^{(N)\dagger}_{\pi_{i,j}}(U\otimes \mathbb{I}_{2^{N-2}})Sw^{(N)}_{\pi_{i,j}}
\end{align*}
 Again, we consider the all-zero codeword as the input to the decoder so that all channels are identical i.e $W_{i}(z)=\ketbra{(-1)^{z}\theta}{(-1)^{z}\theta}$, $\forall i\in\{1,2,3,4\}$ and $z\in\{0,1\}$.
 \subsubsection{Decoding bit $u_{1}$}
 If $u_1$ is an information bit, then we decode it based on the decoding factor graph for bit $u_1$ shown in Fig.~\ref{polar4dec1}.
 First we apply check node $U_{1,2}^{(4)}$ unitary on the joint state $\ket{\theta}_{1}\otimes\ket{\theta}_{2}\otimes\ket{\theta}_{3}\otimes\ket{\theta}_{4}$ which produces a new state denoted as $\ket{\eta_{1}}_{A_1A_2A_3A_4}$- 
        \begin{align*}
       &\ket{\eta_{1}}_{A_1A_2A_3A_4}    = U_{1,2}^{(4)}\ket{\theta}_{A_1}\otimes\ket{\theta}_{A_2}\otimes\ket{\theta}_{A_3}\otimes\ket{\theta}_{A_4} \\
            & \quad = \left( \sum_{j\in\{0,1\}}\sqrt{p_{j}}\ket{\theta_{j}^{\cnop}}_{A_1}\otimes\ket{j}_{A_2}\right)
             \otimes \ket{\theta}_{A_3}\otimes\ket{\theta}_{A_4}
        \end{align*}
        where $p_{j}$ with $j\in\{0,1\}$ denotes the probability  of state $\ket{\theta_{j}^{\cnop}}_{A_1}$.
 Similarly, we apply $U_{3,4}^{(4)}$ on state  $\ket{\eta_{1}}_{A_1A_2A_3A_4}$ as
        \begin{align*}
     \ket{\eta_{2}}_{A_1A_2A_3A_4}&    
 =  U_{3,4}^{(4)} \ket{\eta_{1}}_{A_1A_2A_3A_4} \\
           & = \left( \sum_{j\in\{0,1\}}\sqrt{p_{j}}\ket{\theta_{j}^{\cnop}}_{A_1}\otimes \ket{j}_{A_2} \right) \\
           & \quad\quad \otimes\left(\sum_{j\in\{0,1\}}\sqrt{p_{j}}\ket{\theta_{j}^{\cnop}}_{A_3}\otimes \ket{j}_{A_4}\right).
        \end{align*}
         From the factor graph depicted in Fig.~\ref{polar4dec1}, we see that, in order to estimate the input bit $u_{1}$, we need to apply $U_{1,3}^{(4)}$  on $\ket{\eta_{2}}_{A_1A_2A_3A_4}$. In the resulting state $\ket{\eta_{3}}_{A_1A_2A_3A_4} =U_{1,3}\ket{\eta_{2}}_{A_1A_2A_3A_4})$,  all the decision information is compressed to system $A_{1}$. For the pure state case, states $\ket{\theta}$ and $\ket{-\theta}$ are related to each other by the unitary $\sigma_{z}$, so we measure the state of $A_{1}$ in the $\sigma_{x}$ basis by forming the Helstrom measurement $\hat{\Pi}_{1}^{+}=\{\Pi_{1}^{+},\mathbb{I}_{16}-\Pi_{1}^{+}\}$ where $\Pi_{1}^{+}=\ketbra{+}{+}\otimes \mathbb{I}_{8}$. The error probability for estimating the bit $u_{1}$ thus becomes
        \begin{align}
p_{u_{1}}=1-\bra{\eta_{3}}\Pi_{1}^{+}\ket{\eta_{3}}_{A_1A_2A_3A_4}.
        \end{align}

\subsubsection{Decoding $u_{2}$}
     If $u_2$ is an information bit, then we decode it based on the decoding factor graph for bit $u_2$ depicted Fig.~\ref{polar4dec1}. To decode bit $u_{2}$, we use the decoder output $\hat{u}_{1}$ assuming  $\hat{u}_{1}$ is the correct estimate of $u_{1}$. First, to incorporate the output $\hat{u}_{1}$, we apply $\sigma_{z}^{\hat{u}_{1}}\otimes \mathbb{I}_{8}$ on $\ket{\eta_{3}}_{A_1A_2A_3A_4}$. Next,
     we apply  $(U_{1,3}^{(4)})^{\dagger}$  on $\ket{\eta_{3}}_{A_1A_2A_3A_4}$  to get the resulting state $\ket{\eta_{4}}_{A_1A_2A_3A_4}$. 
        Next, we apply swap operator $Sw_{\pi'}^{(4)}$ on $\ket{\eta_{4}}_{A_1A_2A_3A_4}$. Similar to the general qubit BSCQ scenario, we design the conditional bit node unitary $V_{3,4|1,2}^{cond}(\{\theta_{0}^{\cnop},\theta_{1}^{\cnop}\},\{\theta_{0}^{\cnop},\theta_{1}^{\cnop}\})$ such that conditioned on the first two qubits we can apply bit node unitaries $V$ coherently on the qubits 3 and 4. The unitary $V_{3,4|1,2}^{cond}$ takes the following form 
 \begin{align}
 &\ V_{3,4|1,2}^{cond}(\{\theta_{0}^{\cnop},\theta_{1}^{\cnop}\},\{\theta_{0}^{\cnop},\theta_{1}^{\cnop}\}) \nonumber\\
 &\ =
  \begin{bmatrix}
   V(\theta_{0}^{\cnop},\theta_{0}^{\cnop})& &&\\
   & V(\theta_{0}^{\cnop},\theta_{1}^{\cnop})& & \\
   & & V(\theta_{1}^{\cnop},\theta_{0}^{\cnop})& \\
   & & & V(\theta_{1}^{\cnop},\theta_{1}^{\cnop})
  \end{bmatrix}.
 \end{align}
 
 We apply $V_{3,4|1,2}^{cond}(\{\theta_{0}^{\cnop},\theta_{1}^{\cnop}\},\{\theta_{0}^{\cnop},\theta_{1}^{\cnop}\})$ to get the state $\ket{\eta_{5}}_{A_1A_2A_3A_4}$ and measure the state corresponding to system $A_{3}$ by forming the Helstrom measurement $\hat{\Pi}_{3}^{+}=\{\Pi_{3}^{+},\mathbb{I}_{16}-\Pi_{3}^{+}\}$ where $\Pi_{3}^{+}=\mathbb{I}_{4}\otimes \ketbra{+}{+}\otimes \mathbb{I}_{2}$. Then the error probability of estimating bit $u_{2}$ is
 \begin{align}
p_{u_{2}}=1-\bra{\eta_{5}}\Pi^{+}_{3}\ket{\eta_{5}}_{A_1A_2A_3A_4}.
 \end{align}

\subsubsection{Decoding $u_{3}$}
     
If $u_3$ is an information bit, then we decode it based on the decoding factor graph for bit $u_3$ depicted in Fig.~\ref{polar4dec3}.
First, we  apply $\sigma_{z}^{\hat{u}_{2}}\otimes \mathbb{I}_{8}$ on $\ket{\eta_{5}}_{A_1A_2A_3A_4}$. On the resulting state we apply $\Big{(}V_{3,4|1,2}^{cond}(\{\theta_{0}^{\cnop},\theta_{1}^{\cnop}\},\{\theta_{0}^{\cnop},\theta_{1}^{\cnop}\})\Big{)}^{\dagger}$  and then  $(Sw_{\pi'}^{(4)})^{\dagger}$ to get new state $\ket{\eta_{6}}_{A_1A_2A_3A_4}$. On $\ket{\eta_{6}}_{A_1A_2A_3A_4}$ we apply $(U_{3,4}^{(4)})^{\dagger}$ and $(U_{1,2}^{(4)})^{\dagger}$  to produce a new state $\ket{\eta_{7}}_{A_1A_2A_3A_4}$. Next, we apply bit node unitary $V_{1,2}^{(4)}(\theta,\theta)$ and the bit node unitary $V_{3,4}^{(4)}(\theta,\theta)$ on $\ket{\eta_{7}}_{A_1A_2A_3A_4}$. Finally, we apply $U_{1,3}^{(4)}$ on the resulting state to get a new state $\ket{\eta_{8}}_{A_1A_2A_3A_4}$ which we measure by forming the measurement $\hat{\Pi}_{1}^{+}$. Thus, the error
probability  for estimating bit $u_{3}$ becomes 
\begin{align}
    p_{u_{3}}=1-\bra{\eta_{8}}\Pi^{+}_{1}\ket{\eta_{8}}_{A_1A_2A_3A_4}.
\end{align}

\subsubsection{Decoding $u_{4}$}
    If $u_4$ is an information bit, then we decode it based on the decoding factor graph for bit $u_4$ depicted in Fig.~\ref{polar4dec3}.
To decode  $u_{4}$, first we apply $\sigma_{z}^{\hat{u}_{3}}\otimes \mathbb{I}_{8}$ on $\ket{\eta_{8}}_{A_1A_2A_3A_4}$
which is followed by  $(U_{1,3}^{(4)})^{\dagger}$. Finally,  we apply
          $V_{1,3}^{(4)}(\theta\vnop\theta,\theta\vnop\theta)$ to get the resulting state denoted as $\ket{\eta_{9}}_{A_1A_2A_3A_4}$ which we measure using $\hat{\Pi}_{1}^{+}$. Hence, the error
probability  for estimating bit $u_{4}$ becomes 
\begin{align}
    p_{u_{4}}=1-\bra{\eta_{9}}\Pi^{+}_{1}\ket{\eta_{9}}_{A_1A_2A_3A_4}.
\end{align}

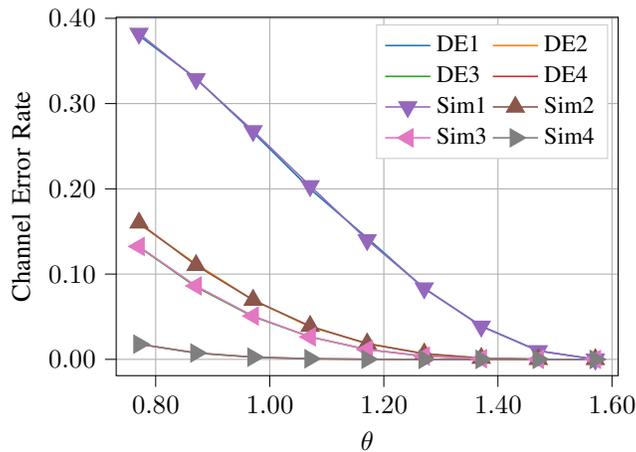
\begin{figure}[ht]
    \centering
\begin{tikzpicture}

\definecolor{crimson2143940}{RGB}{214,39,40}
\definecolor{darkgray176}{RGB}{176,176,176}
\definecolor{darkorange25512714}{RGB}{255,127,14}
\definecolor{forestgreen4416044}{RGB}{44,160,44}
\definecolor{gray127}{RGB}{127,127,127}
\definecolor{lightgray204}{RGB}{204,204,204}
\definecolor{mediumpurple148103189}{RGB}{148,103,189}
\definecolor{orchid227119194}{RGB}{227,119,194}
\definecolor{sienna1408675}{RGB}{140,86,75}
\definecolor{steelblue31119180}{RGB}{31,119,180}

\begin{axis}[
width=3.25in,
height=2.5in,
legend cell align={left},
legend style={fill opacity=0.8, draw opacity=1, text opacity=1, draw=lightgray204,font=\small, legend columns=2},
tick align=outside,
tick pos=left,
x grid style={darkgray176},
xlabel={\(\displaystyle \theta\)},
xmajorgrids,
xmin=0.730796326794896, xmax=1.6107963267949,
xminorgrids,
xtick style={color=black},
xticklabel style={
            /pgf/number format/precision=2,
            /pgf/number format/fixed, /pgf/number format/fixed zerofill
        },
y grid style={darkgray176},
ylabel={Channel Error Rate},
ymajorgrids,
ymin=-0.0191096652031247, ymax=0.401302969265619,
yminorgrids,
ytick style={color=black},
yticklabel style={
            /pgf/number format/precision=2,
            /pgf/number format/fixed, /pgf/number format/fixed zerofill
        },
]
\addplot [semithick, steelblue31119180, mark=Mercedes star flipped*, mark size=3.75, mark options={solid}]
table {%
1.5707963267949 0
1.4707963267949 0.0100478661650293
1.3707963267949 0.0390761885366693
1.2707963267949 0.0831169208909454
1.1707963267949 0.14237566540265
1.0707963267949 0.199817585597212
0.970796326794896 0.266289934504419
0.870796326794896 0.329452954373016
0.770796326794897 0.379806507517531
};
\addlegendentry{DE1}
\addplot [semithick, darkorange25512714, mark=Mercedes star*, mark size=3.75, mark options={solid}]
table {%
1.5707963267949 0
1.4707963267949 9.94051145394304e-05
1.3707963267949 0.00149826464957381
1.2707963267949 0.006643509138755
1.1707963267949 0.0187165845023176
1.0707963267949 0.0381214788376373
0.970796326794896 0.0696182413419772
0.870796326794896 0.111808851855008
0.770796326794897 0.158966896837877
};
\addlegendentry{DE2}
\addplot [semithick, forestgreen4416044, mark=Mercedes star*, mark size=3.75, mark options={solid,rotate=90}]
table {%
1.5707963267949 0
1.4707963267949 5.01394938294819e-05
1.3707963267949 0.000775220615833017
1.2707963267949 0.00376112671866502
1.1707963267949 0.0115006375288197
1.0707963267949 0.0263492478686181
0.970796326794896 0.0501292790572831
0.870796326794896 0.0849263354184328
0.770796326794897 0.131839808233942
};
\addlegendentry{DE3}
\addplot [semithick, crimson2143940, mark=Mercedes star*, mark size=3.75, mark options={solid,rotate=270}]
table {%
1.5707963267949 0
1.4707963267949 2.46687692495584e-09
1.3707963267949 6.06718032925002e-07
1.2707963267949 1.4542657598926e-05
1.1707963267949 0.000132229595325772
1.0707963267949 0.000698252791343723
0.970796326794896 0.00258973019518705
0.870796326794896 0.00747237329250516
0.770796326794897 0.0178500957986574
};
\addlegendentry{DE4}
\addplot [semithick, mediumpurple148103189, mark=triangle*, mark size=3.75, mark options={solid,rotate=180}]
table {%
1.5707963267949 2.22044604925031e-16
1.4707963267949 0.00991704341450905
1.3707963267949 0.0386905821650815
1.2707963267949 0.0835187366177882
1.1707963267949 0.140148292807039
1.0707963267949 0.203433600817162
0.970796326794896 0.267997668602159
0.870796326794896 0.328897110566731
0.770796326794897 0.382193304062495
};
\addlegendentry{Sim1}
\addplot [semithick, sienna1408675, mark=triangle*, mark size=3.75, mark options={solid}]
table {%
1.5707963267949 3.33066907387547e-16
1.4707963267949 9.74216689968443e-05
1.3707963267949 0.00144956145310871
1.2707963267949 0.00659825523851221
1.1707963267949 0.0183552342346497
1.0707963267949 0.0389382842098346
0.970796326794896 0.0695738060650879
0.870796326794896 0.110346843819858
0.770796326794897 0.160168674602223
};
\addlegendentry{Sim2}
\addplot [semithick, orchid227119194, mark=triangle*, mark size=3.75, mark options={solid,rotate=90}]
table {%
1.5707963267949 1.11022302462516e-15
1.4707963267949 4.96676648700234e-05
1.3707963267949 0.000778920833477192
1.2707963267949 0.00381345592737248
1.1707963267949 0.0114983525193786
1.0707963267949 0.026415246248769
0.970796326794896 0.0508234541595034
0.870796326794896 0.0861193179831495
0.770796326794897 0.132406457088151
};
\addlegendentry{Sim3}
\addplot [semithick, gray127, mark=triangle*, mark size=3.75, mark options={solid,rotate=270}]
table {%
1.5707963267949 1.33226762955019e-15
1.4707963267949 2.46687714700045e-09
1.3707963267949 6.06718033480114e-07
1.2707963267949 1.4542657598704e-05
1.1707963267949 0.000132229595326105
1.0707963267949 0.000698252791343945
0.970796326794896 0.00258973019518671
0.870796326794896 0.00747237329250616
0.770796326794897 0.0178500957986584
};
\addlegendentry{Sim4}
\end{axis}

\end{tikzpicture}
    \vspace{-1mm}
    \caption{Comparison of bit-error rate between DE analysis and simulated decoder for bits $u_1,\ldots,u_4$ of length-4 polar code over the PSC with parameter $\theta$.}
    \label{devsdecoder-psc}
\end{figure}

\subsection{PM-BPQM Polar Decoder}
To describe the length $N=2^n$ PM-BPQM polar decoder, we start  with defining the necessary tools. As discussed in the earlier sections, the bit-node and check-node unitaries are used to compress the decision information into the first qubit while retaining the reliability information in the second qubit.
To apply the next set of bit-node and check-node unitary on the information qubits according to the certain factor graph we need to know BSCQ channel parameters i.e. $\delta$ and $\gamma$ parameters corresponding to reliability qubits.
The reliability of the effective channel associated with a qubit may depend on other qubit values and operations that depend on the channel $(\delta,\gamma)$ parameters can be implemented as conditional unitaries that depend on these other qubits.

To describe these operations, let us define the conditional list $L(i) = (L_1 (i),L_2 (i),\ldots,L_{l_i} (i))$ for system $A_i$ that contains the indices of the $l_i$ reliability qubits that determine the reliability of the information qubit 
Similarly, the conditional parameters $\delta^{(i)}=(\delta_{0}^{(i)},\ldots,\delta_{2^{l_i}-1}^{(i)})$ and $\gamma^{(i)}=(\gamma_{0}^{(i)},\ldots,\gamma_{2^{l_i}-1}^{(i)})$ define the parameters $(\delta^{(i)}_m,\gamma^{(i)}_m)$ for the effective channel associated with the $i$-th qubit assuming the reliability qubits in $L(i)$ are in the $m$-th computational basis state where $m\in \{0,\ldots,2^{l_i}-1\}$. 
Whenever we apply bit-node and check-node operations on the information qubits, we get two post-measurement states as outputs \cite{brandsen_bpqm_arxiv} which are characterized by ($\delta,\gamma$) parameters.
This process is responsible for the dependence of the channel parameters on reliability qubits.
Next, we define a permutation $\pi_{i,j,L(i),L(j)}$ such that 
 \begin{align}
     \pi_{i,j,L(i),L(j)}(k)=\begin{cases}
     m,  \quad \text{if $k=L_{m}(i)$}\\
     m+l_i, \quad \text{if $k=L_m(j)$}\\
     n-1, \quad \text{if $k=i$}\\
     n, \qquad \text{if $k=j$}
     \end{cases}
 \end{align}
 With these tools, we can generalize the conditional bit-node unitary.
 Let $V^{cond}_{i,j|L(i),L(j)}\left ( \delta^{(i)},\gamma^{(i)},\delta^{(j)},\gamma^{(j)}\right)$ be the conditional bit-node unitary operating on information qubits in systems $A_{i}$ and $A_{j}$ with reliability qubits specified using $L(i)$ and $L(j)$.
 Then, $Sw^{(N),\dagger}_{\pi_{i,j,L(i),L(j)}}V^{cond}_{i,j|L(i),L(j)}\left (\delta^{(i)},\gamma^{(i)},\delta^{(j)},\gamma^{(j)}\right)Sw^{(N)}_{\pi_{i,j,L(i),L(j)}}$ can be written as
 \begin{align*}
&\ Sw^{(N)}_{\pi_{i,j,L(i),L(j)}}V^{cond}_{i,j|L(i),L(j)}\left (\delta^{(i)},\gamma^{(i)},\delta^{(j)},\gamma^{(j)}\right)Sw^{(N),\dagger}_{\pi_{i,j,L(i),L(j)}}\\
&\ \qquad = \underbrace{\begin{bmatrix}
    V(L(i),L(j))& & & \\
    & V(L(i),L(j)) & & \\
& & \ddots &\\
&&& V(L(i),L(j))
\end{bmatrix}}_{\text{$2^{n-l_i-l_j-2}$ blocks}},
\end{align*}
  where $V(L(i),L(j))$ is block diagonal with $2^{l_i+l_j}$ 4 by 4 blocks and block $k \in \{0,\ldots,2^{l_i+l_j}-1\}$ is defined by
  \[ V\left(\delta^{(i)}_{\lfloor{k 2^{-l_j} \rfloor}},\gamma^{(i)}_{\lfloor{k 2^{-l_j} \rfloor}},\delta^{(j)}_{k \bmod 2^{l_j}},\gamma^{(j)}_{k \bmod 2^{l_j}} \right) \]
  in terms of the the bit-node unitary $V(\delta,\gamma,\delta',\gamma')$ from Lemma~\ref{bscq-bitnode}.
Thus, we can write
\begin{align*}
V(L(i),L(j))&= \text{diag}\Big( V(\delta_{0}^{(i)},\gamma_{0}^{(i)},\delta_{0}^{(j)},\gamma_{0}^{(j)}), \ldots, V(\delta_{0}^{(i)},\gamma_{0}^{(i)},\delta_{2^{l_j}-1}^{(j)},\gamma_{2^{l_j}-1}^{(j)}), \ldots, \\ & \qquad \qquad  \quad V(\delta_{2^{l_i}-1}^{(i)},\gamma_{2^{l_i}-1}^{(i)},\delta_{0}^{(j)},\gamma_{0}^{(j)}), \ldots, V(\delta_{2^{l_i}-1}^{(i)},\gamma_{2^{l_i}-1}^{(i)},\delta_{2^{l_j}-1}^{(j)},\gamma_{2^{l_j}-1}^{(j)}) \Big).
\end{align*}
 In Algorithm~\ref{polar_cq_vnop}, we discuss the pseudo-code to apply the bit-node operation using PM-BPQM on the information qubits with given a conditional list and a conditional parameter list for the quantum state $\ket{\psi}_{A_1\dots A_{N}}$. We later refer to this pseudo-code as \textbf{BNOP} when we describe the PM-BPQM polar decoder over BSCQ channels.
 \begin{algorithm}[b]
     \caption{Bit-node operation using PM-BPQM (BNOP)\label{polar_cq_vnop}}
     \begin{algorithmic}[1]
     \Require A quantum state $\ket{\psi}_{A_1\dots A_{N}}$; indices $(i,j)$ of quantum systems $A_i$ and $A_j$ for the information qubits; conditional lists of reliability qubits $L(i)$ and $L(j)$ corresponding to information qubits; conditional parameter lists $\delta^{(i)},\gamma^{(i)}$ and $\delta^{(j)},\gamma^{(j)}$ corresponding to information qubits. 
     \State Create a new conditional list $L'(i)$ for the information qubit in $A_{i}$ such that $L'(i)=\{j\}\bigcup L(i)\bigcup L(j)$.\\
     Initialize new empty conditional parameter lists $\delta^{(i)\prime} ,\gamma^{(i)\prime}$  for the updated information qubit in $A_{i}$.\\
     Apply permutation $\pi_{i,j,L(i),L(j)}$ using swap operator on the quantum state :
$\ket{\psi}_{A_1\dots A_{N}}\gets Sw^{(N)}_{\pi_{i,j,L(i),L(j)}}\ket{\psi}_{A_1\dots A_{N}} $.\\
Apply conditional bit-node unitary $V^{cond}_{i,j|L(i),L(j)}\left (\delta^{(i)},\gamma^{(i)},\delta^{(j)},\gamma^{(j)}\right)$ followed by swap operator $Sw^{(N)}_{\pi_{i,j,L(i),L(j)}}$ on the quantum state:
$\ket{\psi}_{A_1\dots A_{N}}\gets Sw^{(N)}_{\pi_{i,j,L(i),L(j)}}V^{cond}_{i,j|L(i),L(j)}\left (\delta^{(i)},\gamma^{(i)},\delta^{(j)},\gamma^{(j)}\right)Sw^{(N),\dagger}_{\pi_{i,j,L(i),L(j)}}\ket{\psi}_{A_1\dots A_{N}} $.\\
Apply inverse of permutation $\pi_{i,j,L(i),L(j)}$ on the quantum state :
$\ket{\psi}_{A_1\dots A_{N}}\gets Sw^{(N),\dagger}_{\pi_{i,j,L(i),L(j)}}\ket{\psi}_{A_1\dots A_{N}} $.
\For{each element $(\delta_{o},\gamma_{o})  \in (\delta^{(i)},\gamma^{(i)})$ }
 \For{each element $(\delta_{q},\gamma_{q})  \in (\delta^{(j)},\gamma^{(j)})$ }\\
\qquad Use PM-BPQM to compute the bit-node parameters $\{\delta_{a}^{\vnop} (\delta_{o},\gamma_{o},\delta_{q},\gamma_{q}),\gamma_{a}^{\vnop} (\delta_{o},\gamma_{o},\delta_{q},\gamma_{q})\}_{a\in \{0,1\}}$\\
\qquad Add $\{\delta_{a}^{\vnop} (\delta_{o},\gamma_{o},\delta_{q},\gamma_{q})\}_{a\in \{0,1\}}$ to the list $\delta^{(i)\prime}$ and add $\gamma_{a}^{\vnop} (\delta_{o},\gamma_{o},\delta_{q},\gamma_{q})\}_{a\in \{0,1\}}$ to the list $\gamma^{(i)\prime}$.
\EndFor
\EndFor\\
\Return Quantum state $\ket{\psi}_{A_1\dots A_{N}}$, new conditional list $L(i)$, and new conditional parameter lists $\delta^{(i)\prime},\gamma^{(i)\prime}$.
     \end{algorithmic}
\end{algorithm}

Since the check-node unitary does not depend on the channel parameters $(\delta,\gamma)$, the unitary $C^{(N)}_{i,j}$ defined $\forall i,j\in [N]$ is sufficient to realize the complete polar decoder. In Algorithm \ref{polar_cq_cnop}, we define the check-node operation using PM-BPQM on the information qubits with given a conditional list of parameters for the quantum state $\ket{\psi}_{A_1\dots A_{N}}$.
We later refer to this pseudo-code as \textbf{CNOP} when we describe the PM-BPQM polar decoder over BSCQ channels. Both bit-node and check-node operations are invertible and can be implemented by using conjugate transpose of all unitary corresponding to every operation. We denote these operations as \textbf{BNOP}$^{-1}$ and \textbf{CNOP}$^{-1}$.

 \begin{algorithm}[t]
     \caption{Check-node operation using PM-BPQM (CNOP)}\label{polar_cq_cnop}
     \begin{algorithmic}[1]
     \Require A quantum state $\ket{\psi}_{A_1\dots A_{N}}$; indices $(i,j)$ of quantum systems $A_i$ and $A_j$ for the information qubits; conditional lists of reliability qubits $L(i)$ and $L(j)$ corresponding to information qubits; conditional parameter lists $\delta^{(i)},\gamma^{(i)}$ and $\delta^{(j)},\gamma^{(j)}$ corresponding to the reliability of qubits $L(i)$ and $L(j)$.  
     \State Create a new conditional list $L'(i)$ for the information qubit in $A_{i}$ such that $L'(i)=\{j\}\bigcup L(i)\bigcup L(j)$.\\
     Initialize new empty conditional parameter lists $\delta^{(i)\prime} ,\gamma^{(i)\prime}$  for the updated information qubit in $A_{i}$.\\
     Apply permutation $\pi_{i,j}$ using swap operator on the quantum state :
$\ket{\psi}_{A_1\dots A_{N}}\gets Sw^{(N)}_{\pi_{i,j}}\ket{\psi}_{A_1\dots A_{N}} $.\\
Apply check-node unitary $C^{(N)}_{i,j}$ on the quantum state:
$\ket{\psi}_{A_1\dots A_{N}}\gets C^{(N)}_{i,j}\ket{\psi}_{A_1\dots A_{N}} $.\\
Apply inverse of permutation $\pi_{i,j}$ on the quantum state :
$\ket{\psi}_{A_1\dots A_{N}}\gets Sw^{(N),\dagger}_{\pi_{i,j}}\ket{\psi}_{A_1\dots A_{N}} $.
\For{each element $(\delta_{o},\gamma_{o})  \in (\delta^{(i)},\gamma^{(i)})$ }
 \For{each element  $(\delta_{q},\gamma_{q})  \in  (\delta^{(j)},\gamma^{(j)})$ }\\
\qquad Use PM-BPQM compute to bit-node parameters $\{\delta_{a}^{\cnop} (\delta_{o},\gamma_{o},\delta_{q},\gamma_{q}),\gamma_{a}^{\cnop} (\delta_{o},\gamma_{o},\delta_{q},\gamma_{q})\}_{a\in \{0,1\}}$\\
\qquad Add $\{\delta_{a}^{\cnop} (\delta_{o},\gamma_{o},\delta_{q},\gamma_{q})\}_{a\in \{0,1\}}$ to the list $\delta^{(i)\prime}$ and add $\gamma_{a}^{\cnop} (\delta_{o},\gamma_{o},\delta_{q},\gamma_{q})\}_{a\in \{0,1\}}$ to the list $\gamma^{(i)\prime}$.
\EndFor
\EndFor\\
\Return Quantum state $\ket{\psi}_{A_1\dots A_{N}}$, new conditional list $L(i)$, and new conditional parameter lists $\delta^{(i)\prime},\gamma^{(i)\prime}$.
     \end{algorithmic}
 \end{algorithm}
 
 To describe the PMBPQM-Polar decoder first, we denote $\mathcal{I}$ as the index set containing indices of the code bits and $\mathcal{I}(m)$ denotes the $m$-th element in $\mathcal{I}$. For index set $\mathcal{I}$, we define conditional list $L(\mathcal{I})$ which contains conditional lists for all the elements in $\mathcal{I}$ i.e. $L(\mathcal{I})=\{L(m)\}_{m\in \mathcal{I}}$. Similarly we define conditional parameter list $L_{(\delta,\gamma)}(\mathcal{I})$ for $\mathcal{I}$ which contains conditional parameter lists for all the elements in $\mathcal{I}$ i.e. $L_{(\delta,\gamma)}(\mathcal{I})=\{L_{\delta,\gamma}(m)\}_{m\in \mathcal{I}}$. The decoder uses the knowledge of frozen set $\mathcal{A}^c$. For index set $\mathcal{I}$, we define corresponding information index set $\mathcal{I}_{\mathcal{A}}$ such that 
 \begin{align*}
     \mathcal{I}_{\mathcal{A}}(m)=\begin{cases}
         1 ,\quad \text{if $m\in \mathcal{A}$}\\
         0 ,\quad \text{otherwise}.
     \end{cases}
 \end{align*}
 We consider $F(\mathcal{I}_{\mathcal{A}})$ to be the frozen value set which contains the values of frozen bits. In Algorithm \ref{polar_cq} we describe the PM-BPQM-based  
 decoding algorithm for polar codes over BSCQ channels which we call as \textbf{Polar-BSCQ}. The input bits $u_{1}^{N}$ are mapped to the encoded  sequence $x_{1}^{N}=u_{\mathcal{A}}G_{N}(\mathcal{A})\oplus u_{\mathcal{A}^{c}}G_{N}(\mathcal{A}^{c})$  where $\mathcal{A}$ is the information set and $G_{N}(\mathcal{A})$ denotes submatrix of $G_{N}$ constructed from the rows of $G_{N}$ with indices in $\mathcal{A}$. The sequence $x_{1}^{N}$ is transmitted over the BSCQ channel which is characterized by $W^{A_1...A_{N}}(u_{1}^{N})$ such that 
 \begin{align*}
     W^{A_1...A_{N}}(u_{1}^{N})=\prod_{i=1}^{N}W^{A_{i}}(x_i).
 \end{align*}
 A sampled quantum state $\ket{\psi}_{A_1..A_{N}}$ from the output state $ W^{A_1...A_{N}}(u_{1}^{N})$ is passed through the decoder and we analyze the performance of the decoder. 
 
\begin{algorithm}
     \caption{PM-BPQM Polar Decoder over BSCQ channel (Polar-BSCQ)}\label{polar_cq}
     \begin{algorithmic}[1]
     \Require A quantum state $\ket{\psi}_{A_1\dots A_{N}}$, index set $\mathcal{I}$ containing indices of code bits such that $|\mathcal{I}|=N_\mathcal{I}$; conditional list $L(\mathcal{I})$ for the index set $\mathcal{I}$; conditional parameter list $\delta^{(\mathcal{I})},\gamma^{(\mathcal{I})}$ for the index set $\mathcal{I}$; code design vector $F \in \{I,0,1\}^{N_{\mathcal{I}}}$ (either information bit I or frozen to 0 or 1). 
\If{$N_\mathcal{I}=1$} \Comment{Recurse down to length 1}
 \If{$F(1) = I$} \Comment{If the bit is information bit}
  \qquad \State measure the first qubit in the index set $\mathcal{I}$ to set the hard decision output $\hat{u}_{1}$.
  \qquad \If{$\hat{u}_{1}=1$}
  \qquad \State flip the qubit corresponding to the system $\mathcal{I}(1)$ by applying $\sigma_{x}$.
  \EndIf\\
  \qquad \Return Hard decision $\hat{u}_{1}$, post-measurement quantum state $\ket{\psi}_{A_1\dots A_{N}}$
  \Else
  \State $\hat{u}_{1} = F(1)$
\qquad \If{$\hat{u}_{1}=1$} \Comment{If frozen bit is 1}
  \qquad \State flip the qubit corresponding to the system $\mathcal{I}(1)$ by applying $\sigma_{x}$.
  \EndIf\\
   \qquad   \Return Bit value $\hat{u}_{1}$, quantum state $\ket{\psi}_{A_1\dots A_{N}}$
 \EndIf
\EndIf\\
Define $\mathcal{I}_{odd} = \{ \mathcal{I}(2i-1) \, | \, i\in [N_{\mathcal{I}}/2] \}$.\\
Create an empty conditional list $L'$ and parameter lists $\delta^{(i)\prime},\gamma^{(i)\prime}$ for $i \in \mathcal{I}_{odd}$
\For{each odd $i$ in $[N_\mathcal{I}]$}\\
\quad $\ket{\psi}_{A_1\dots A_{N}},L'(\mathcal{I}(i)),\delta^{(\mathcal{I}(i))\prime},\gamma^{(\mathcal{I}(i))\prime}\gets$ CNOP$\left (\ket{\psi}_{A_1\dots A_{N}},L(\mathcal{I}(i)),L(\mathcal{I}(i+1)),\delta^{(\mathcal{I}(i))},\gamma^{(\mathcal{I}(i))},\delta^{(\mathcal{I}(i+1))},\gamma^{(\mathcal{I}(i+1))}\right )$.
\EndFor\\
$\hat{u}_{1}^{N_\mathcal{I} / 2},\ket{\psi}_{A_1\dots A_{N}} \gets $ Polar-BSCQ$\left (\ket{\psi}_{A_1\dots A_{N}}, \mathcal{I}_{odd},L'(\mathcal{I}_{odd}),\delta^{(\mathcal{I}_{odd})\prime},\gamma^{(\mathcal{I}_{odd})\prime},F([N_\mathcal{I}/2])\right )$
\For{each odd $i$ in $[N_\mathcal{I}]$}\\
\quad $\ket{\psi}_{A_1\dots A_{N}}\gets$ CNOP$^{-1}\left (\ket{\psi}_{A_1\dots A_{N}},L(\mathcal{I}(i)),L(\mathcal{I}(i+1)),\delta^{(\mathcal{I}(i))},\gamma^{(\mathcal{I}(i))},\delta^{(\mathcal{I}(i+1))},\gamma^{(\mathcal{I}(i+1))}\right )$.
\EndFor\\
Create an empty conditional list $L'$ and empty conditional parameter lists $\delta^{(i)\prime},\gamma^{(i)\prime}$ for $i \in \mathcal{I}_{odd}$.
\For{each odd $i$ in $[N_\mathcal{I}]$}\\
\quad $\ket{\psi}_{A_1\dots A_{N}},L'(\mathcal{I}(i)),\delta^{(\mathcal{I}(i))\prime},\gamma^{(\mathcal{I}(i))\prime}\gets$ BNOP$\left (\ket{\psi}_{A_1\dots A_{N}},L(\mathcal{I}(i)),L(\mathcal{I}(i+1)),\delta^{(\mathcal{I}(i))},\gamma^{(\mathcal{I}(i))},\delta^{(\mathcal{I}(i+1))},\gamma^{(\mathcal{I}(i+1))}\right )$.
\EndFor\\
$\hat{u}_{N_\mathcal{I}/2}^{N_\mathcal{I}},\ket{\psi}_{A_1\dots A_{N}} \gets $ Polar-BSCQ$\left (\ket{\psi}_{A_1\dots A_{N}},  \mathcal{I}_{odd},L'(\mathcal{I}_{odd}),\delta^{(\mathcal{I}_{odd})\prime},\gamma^{(\mathcal{I}_{odd})\prime},F(\{N_\mathcal{I}/2+1,\ldots,N_\mathcal{I}\})\right )$
\For{each odd element $i$ in $[N_\mathcal{I}]$ }\\
\quad $\ket{\psi}_{A_1\dots A_{N}}\gets$ BNOP$^{-1}\left (\ket{\psi}_{A_1\dots A_{N}},
L(\mathcal{I}(i)),L(\mathcal{I}(i+1)),\delta^{(\mathcal{I}(i))},\gamma^{(\mathcal{I}(i))},\delta^{(\mathcal{I}(i+1))},\gamma^{(\mathcal{I}(i+1))}\right )$.
\EndFor\\
\Return Bit estimates $\hat{u}_{1}^{N_\mathcal{I}}$, the quantum state $\ket{\psi}_{A_1\dots A_{N}}$.
     \end{algorithmic}
 \end{algorithm}

\begin{figure*}
\centering
\begin{minipage}[b]{.48\textwidth}
    \centering
    \scalebox{0.9}{
\begin{tikzpicture}

\definecolor{darkgray176}{RGB}{176,176,176}
\definecolor{darkorange25512714}{RGB}{255,127,14}
\definecolor{forestgreen4416044}{RGB}{44,160,44}
\definecolor{lightgray204}{RGB}{204,204,204}
\definecolor{steelblue31119180}{RGB}{31,119,180}

\begin{axis}[
legend cell align={left},
legend style={fill opacity=0.8, draw opacity=1, text opacity=1, draw=lightgray204},
tick align=outside,
tick pos=left,
x grid style={darkgray176},
xlabel={\(\displaystyle \gamma\)},
xmajorgrids,
xmin=-0.01075, xmax=0.22575,
xminorgrids,
xtick style={color=black},
y grid style={darkgray176},
ylabel={Block Error Rate},
ymajorgrids,
ymin=-0.000875, ymax=0.0538194444444444,
yminorgrids,
ytick style={color=black}
]
\addplot [semithick, steelblue31119180, mark=triangle*, mark size=3.75, mark options={solid,rotate=180}]
table {%
0 0.04445
0.05 0.0513333333333333
0.1 0.0508888888888889
0.15 0.0416666666666667
0.18 0.0338888888888889
0.2 0.0233333333333333
0.215 0.0108333333333333
};
\addlegendentry{Rate=$\frac{1}{2}$}
\addplot [semithick, darkorange25512714, mark=triangle*, mark size=3.75, mark options={solid}]
table {%
0 0.0316111111111111
0.05 0.0313888888888889
0.1 0.0307222222222222
0.15 0.0239444444444444
0.18 0.0186111111111111
0.2 0.0113888888888889
0.215 0.0045
};
\addlegendentry{Rate=$\frac{3}{8}$}
\addplot [semithick, forestgreen4416044, mark=triangle*, mark size=3.75, mark options={solid,rotate=90}]
table {%
0 0.0143888888888889
0.05 0.014
0.1 0.0147777777777778
0.15 0.0122777777777778
0.18 0.00994444444444444
0.2 0.00533333333333333
0.215 0.00161111111111111
};
\addlegendentry{Rate=$\frac{1}{4}$}
\end{axis}

\end{tikzpicture}}
    \vspace{-4mm}
    \caption{Block error rate performance for length-8 CQ Polar code over qubit BSCQ channel with $\delta=0.05$}
    \label{length8-bler_d=0.05}
\end{minipage}\quad
\begin{minipage}[b]{.48\textwidth}
    \centering
    \scalebox{0.9}{
\begin{tikzpicture}

\definecolor{darkgray176}{RGB}{176,176,176}
\definecolor{darkorange25512714}{RGB}{255,127,14}
\definecolor{lightgray204}{RGB}{204,204,204}
\definecolor{steelblue31119180}{RGB}{31,119,180}

\begin{axis}[
legend cell align={left},
legend style={
  fill opacity=0.8,
  draw opacity=1,
  text opacity=1,
  at={(0.03,0.03)},
  anchor=south west,
  draw=lightgray204
},
tick align=outside,
tick pos=left,
x grid style={darkgray176},
xlabel={\(\displaystyle \gamma\)},
xmajorgrids,
xmin=-0.01485, xmax=0.31185,
xminorgrids,
xtick style={color=black},
y grid style={darkgray176},
ylabel={Block Error Rate},
ymajorgrids,
ymin=0.00995555555555556, ymax=0.0622666666666667,
yminorgrids,
ytick style={color=black}
]
\addplot [semithick, steelblue31119180, mark=triangle*, mark size=3.75, mark options={solid,rotate=180}]
table {%
0 0.0545
0.05 0.0554444444444444
0.1 0.0587222222222222
0.15 0.0546666666666667
0.2 0.0480555555555556
0.235 0.0405
0.25 0.035
0.28 0.024
0.297 0.0151111111111111
};
\addlegendentry{Rate=$\frac{1}{4}$ PMBPQM}
\addplot [semithick, darkorange25512714, mark=triangle*, mark size=3.75, mark options={solid}]
table {%
0 0.0533888888888889
0.05 0.0598888888888889
0.1 0.0545555555555556
0.15 0.0407777777777778
0.2 0.0308888888888889
0.235 0.0256111111111111
0.25 0.0218888888888889
0.28 0.0153888888888889
0.297 0.0123333333333333
};
\addlegendentry{Rate=$\frac{1}{4}$ Helstrom}
\end{axis}

\end{tikzpicture}}
    \vspace{-4mm}
    \caption{Block error rate comparison between the PM-BPQM and the successive bitwise Helstrom decoder for length-8 rate $\frac{1}{4}$ CQ polar code over qubit-BSCQ channel with $\delta=0.1$.}
    \label{length8-bler_d=0.1_pmbpqm_helstrom_rate=0.25}
\end{minipage}
\end{figure*}

To analyze the performance of the decoder we choose an information set $\mathcal{A}$
and randomly generate the values of the frozen bits for a sufficiently large number of samples. This gives the average performance of the decoder over the choice of frozen bits. To select the information set $\mathcal{A}$, we rely on the DE results as discussed in Sec.~\ref{sec:pmbpqm_polar_de}. We select $|\mathcal{A}|$ channels with the least channel error rates from the $N$ channels.

 In Fig.~\ref{length8-bler_d=0.05} we plot the block error rate performance of the length-8  PM-BPQM polar decoder where we consider $\mathcal{A}=\{7,8\}$, $\mathcal{A}=\{6,7,8\}$ and $\mathcal{A}=\{4,6,7,8\}$ for the code rates $\frac{1}{4}$, $\frac{3}{8}$ and $\frac{1}{2}$ respectively. In this case, we consider $\delta=0.05$ and vary the $\gamma$ parameter over the admissible range.
 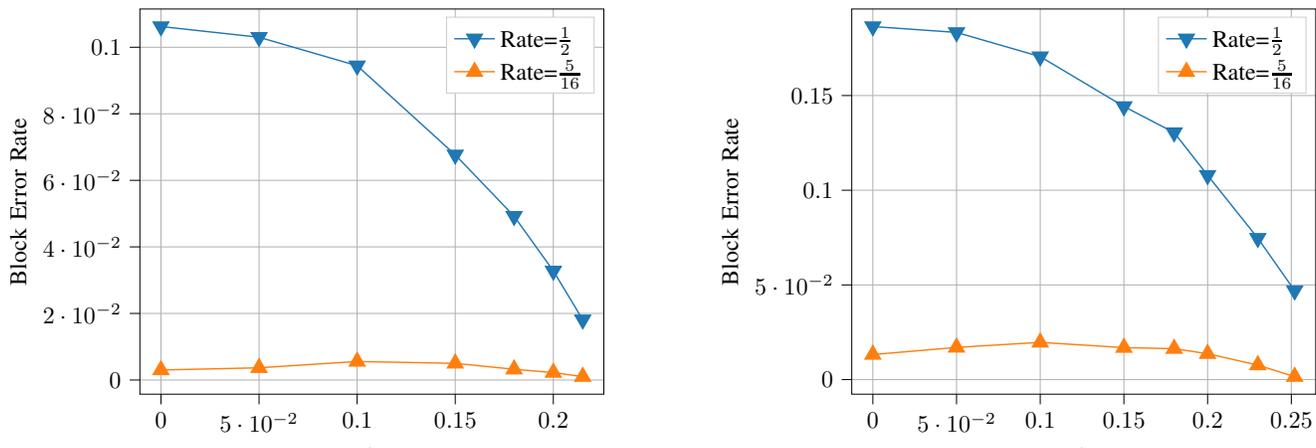
\begin{figure}[t]
    \centering
    \scalebox{0.9}{
\begin{tikzpicture}

\definecolor{darkgray176}{RGB}{176,176,176}
\definecolor{darkorange25512714}{RGB}{255,127,14}
\definecolor{lightgray204}{RGB}{204,204,204}
\definecolor{steelblue31119180}{RGB}{31,119,180}

\begin{axis}[
legend cell align={left},
legend style={fill opacity=0.8, draw opacity=1, text opacity=1, draw=lightgray204},
tick align=outside,
tick pos=left,
x grid style={darkgray176},
xlabel={\(\displaystyle \gamma\)},
xmajorgrids,
xmin=-0.01075, xmax=0.22575,
xminorgrids,
xtick style={color=black},
y grid style={darkgray176},
ylabel={Block Error Rate},
ymajorgrids,
ymin=-0.00426111111111111, ymax=0.111483333333333,
yminorgrids,
ytick style={color=black}
]
\addplot [semithick, steelblue31119180, mark=triangle*, mark size=3.75, mark options={solid,rotate=180}]
table {%
0 0.106222222222222
0.05 0.103
0.1 0.0944444444444444
0.15 0.0676666666666667
0.18 0.0492222222222222
0.2 0.0327777777777778
0.215 0.0181111111111111
};
\addlegendentry{Rate=$\frac{1}{2}$}
\addplot [semithick, darkorange25512714, mark=triangle*, mark size=3.75, mark options={solid}]
table {%
0 0.003
0.05 0.00366666666666667
0.1 0.00555555555555556
0.15 0.005
0.18 0.00322222222222222
0.2 0.00222222222222222
0.215 0.001
};
\addlegendentry{Rate=$\frac{5}{16}$}
\end{axis}

\end{tikzpicture}\hspace{15mm}
\begin{tikzpicture}

\definecolor{darkgray176}{RGB}{176,176,176}
\definecolor{darkorange25512714}{RGB}{255,127,14}
\definecolor{lightgray204}{RGB}{204,204,204}
\definecolor{steelblue31119180}{RGB}{31,119,180}

\begin{axis}[
legend cell align={left},
legend style={fill opacity=0.8, draw opacity=1, text opacity=1, draw=lightgray204},
tick align=outside,
tick pos=left,
x grid style={darkgray176},
xlabel={\(\displaystyle \gamma\)},
xmajorgrids,
xmin=-0.0126, xmax=0.2646,
xminorgrids,
xtick style={color=black},
y grid style={darkgray176},
ylabel={Block Error Rate},
ymajorgrids,
ymin=-0.00768888888888889, ymax=0.195688888888889,
yminorgrids,
ytick style={color=black}
]
\addplot [semithick, steelblue31119180, mark=triangle*, mark size=3.75, mark options={solid,rotate=180}]
table {%
0 0.186444444444444
0.05 0.183333333333333
0.1 0.170555555555556
0.15 0.144222222222222
0.18 0.130444444444444
0.2 0.107888888888889
0.23 0.0746666666666667
0.252 0.047
};
\addlegendentry{Rate=$\frac{1}{2}$}
\addplot [semithick, darkorange25512714, mark=triangle*, mark size=3.75, mark options={solid}]
table {%
0 0.0132222222222222
0.05 0.017
0.1 0.0196666666666667
0.15 0.0168888888888889
0.18 0.0163333333333333
0.2 0.0136666666666667
0.23 0.00766666666666667
0.252 0.00155555555555556
};
\addlegendentry{Rate=$\frac{5}{16}$}
\end{axis}

\end{tikzpicture}}
    \vspace{-4mm}
    \caption{Block error rate performance  for length-16 CQ Polar code over qubit BSCQ channel with $\delta=0.05$ (left) and $\delta=0.07$ (right).}
    \label{length16-bler_d=0.05_0.07}
\end{figure}

In Fig.~\ref{length16-bler_d=0.05_0.07}, we plot the block error rate performance of the length-16  PM-BPQM polar decoder where we consider $\mathcal{A}=\{8,12,14,15,16\}$ and $\mathcal{A}=\{8,10,11,12,13,14,15,16\}$ for the code rates $\frac{5}{16}$ and $\frac{1}{2}$ respectively with channel parameter $\delta=0.05$ and $\delta=0.07$ respectively.


 In Fig.~\ref{length8-bler_d=0.1_pmbpqm_helstrom_rate=0.25}, we compare the block error rate performance of length-8 PM-BPQM polar decoder with the successive bitwise Helstrom decoder for the code rate $\frac{1}{4}$ where we consider channel parameter $\delta=0.1$ and vary parameter $\gamma$. The successive bitwise Helstrom decoder is realized by constructing the measurement as described in \cite{wilde2012polar}. The only difference is that in our case we consider the Helstrom measurement instead of considering the square roots of difference of density matrices. From Fig.~\ref{length8-bler_d=0.1_pmbpqm_helstrom_rate=0.25}, we observe that the PM-BPQM decoder performs better than the successive bitwise Helstrom decoder when $\gamma=0.05$. We should note that this result does not contradict their properties as the successive bitwise Helstrom decoder is not optimal in terms of the block error rate performance.
 
 \begin{figure}[ht]
    \centering
    \scalebox{0.9}{
\begin{tikzpicture}

\definecolor{darkgray176}{RGB}{176,176,176}
\definecolor{darkorange25512714}{RGB}{255,127,14}
\definecolor{forestgreen4416044}{RGB}{44,160,44}
\definecolor{lightgray204}{RGB}{204,204,204}
\definecolor{steelblue31119180}{RGB}{31,119,180}

\begin{axis}[
legend cell align={left},
legend style={
  fill opacity=0.8,
  draw opacity=1,
  text opacity=1,
  at={(0.03,0.03)},
  anchor=south west,
  draw=lightgray204
},
tick align=outside,
tick pos=left,
x grid style={darkgray176},
xlabel={\(\displaystyle \gamma\)},
xmajorgrids,
xmin=-0.01485, xmax=0.31185,
xminorgrids,
xtick style={color=black},
y grid style={darkgray176},
ylabel={Block Error Rate},
ymajorgrids,
ymin=0.0211814573061762, ymax=0.136423710433221,
yminorgrids,
ytick style={color=black}
]
\addplot [semithick, steelblue31119180, mark=triangle*, mark size=3.75, mark options={solid,rotate=180}]
table {%
0 0.111611111111111
0.05 0.113055555555556
0.1 0.123722222222222
0.15 0.103555555555556
0.2 0.0901111111111111
0.235 0.0765
0.25 0.0695
0.28 0.0501666666666667
0.297 0.0353888888888889
};
\addlegendentry{Rate=$\frac{3}{8}$ PMBPQM}
\addplot [semithick, darkorange25512714, mark=triangle*, mark size=3.75, mark options={solid}]
table {%
0 0.113388888888889
0.05 0.125888888888889
0.1 0.114388888888889
0.15 0.0941111111111111
0.2 0.0733888888888889
0.235 0.0627222222222222
0.25 0.0504444444444444
0.28 0.041
0.297 0.0343888888888889
};
\addlegendentry{Rate=$\frac{3}{8}$ Helstrom}
\addplot [semithick, forestgreen4416044, mark=diamond*, mark size=3.75, mark options={solid}]
table {%
0 0.126846791608455
0.05 0.131185426200173
0.1 0.12930272078094
0.15 0.121500013303912
0.2 0.109631298700254
0.235 0.0887092761089791
0.25 0.0770556491763592
0.28 0.047335480781893
0.297 0.0264197415392237
};
\addlegendentry{Rate=$\frac{3}{8}$ UB}
\end{axis}

\end{tikzpicture}\hspace{15mm}
\begin{tikzpicture}

\definecolor{crimson2143940}{RGB}{214,39,40}
\definecolor{darkgray176}{RGB}{176,176,176}
\definecolor{darkorange25512714}{RGB}{255,127,14}
\definecolor{forestgreen4416044}{RGB}{44,160,44}
\definecolor{lightgray204}{RGB}{204,204,204}
\definecolor{steelblue31119180}{RGB}{31,119,180}

\begin{axis}[
legend cell align={left},
legend style={
  fill opacity=0.8,
  draw opacity=1,
  text opacity=1,
  at={(0.09,0.5)},
  anchor=west,
  draw=lightgray204
},
tick align=outside,
tick pos=left,
x grid style={darkgray176},
xlabel={\(\displaystyle \gamma\)},
xmajorgrids,
xmin=-0.01485, xmax=0.31185,
xminorgrids,
xtick style={color=black},
y grid style={darkgray176},
ylabel={Block Error Rate},
ymajorgrids,
ymin=0.00150364337615015, ymax=0.549657802963767,
yminorgrids,
ytick style={color=black}
]
\addplot [semithick, steelblue31119180, mark=triangle*, mark size=3.75, mark options={solid,rotate=180}]
table {%
0 0.111611111111111
0.05 0.113055555555556
0.1 0.123722222222222
0.15 0.103555555555556
0.2 0.0901111111111111
0.235 0.0765
0.25 0.0695
0.28 0.0501666666666667
0.297 0.0353888888888889
};
\addlegendentry{Rate=$\frac{3}{8}$ PMBPQM}
\addplot [semithick, darkorange25512714, mark=triangle*, mark size=3.75, mark options={solid}]
table {%
0 0.113388888888889
0.05 0.125888888888889
0.1 0.114388888888889
0.15 0.0941111111111111
0.2 0.0733888888888889
0.235 0.0627222222222222
0.25 0.0504444444444444
0.28 0.041
0.297 0.0343888888888889
};
\addlegendentry{Rate=$\frac{3}{8}$ Helstrom}
\addplot [semithick, forestgreen4416044, mark=diamond*, mark size=3.75, mark options={solid}]
table {%
0 0.126846791608455
0.05 0.131185426200173
0.1 0.12930272078094
0.15 0.121500013303912
0.2 0.109631298700254
0.235 0.0887092761089791
0.25 0.0770556491763592
0.28 0.047335480781893
0.297 0.0264197415392237
};
\addlegendentry{Rate=$\frac{3}{8}$ UB}
\addplot [semithick, crimson2143940, mark=*, mark size=3.75, mark options={solid}]
table {%
0 0.507387166433821
0.05 0.524741704800694
0.1 0.51721088312376
0.15 0.486000053215646
0.2 0.438525194801016
0.235 0.354837104435916
0.25 0.308222596705437
0.28 0.189341923127572
0.297 0.105678966156895
};
\addlegendentry{Rate=$\frac{3}{8}$ NCUB}
\end{axis}

\end{tikzpicture}}
    \vspace{-4mm}
    \caption{Block error rate comparison for length-8 rate $\frac{3}{8}$ CQ polar code with union bound and non-commutative union bound over qubit-BSCQ channel with $\delta=0.1$.}
    \label{length8-bler_d=0.1_pmbpqm_helstrom_rate_3/8_ub_ncub}
\end{figure}
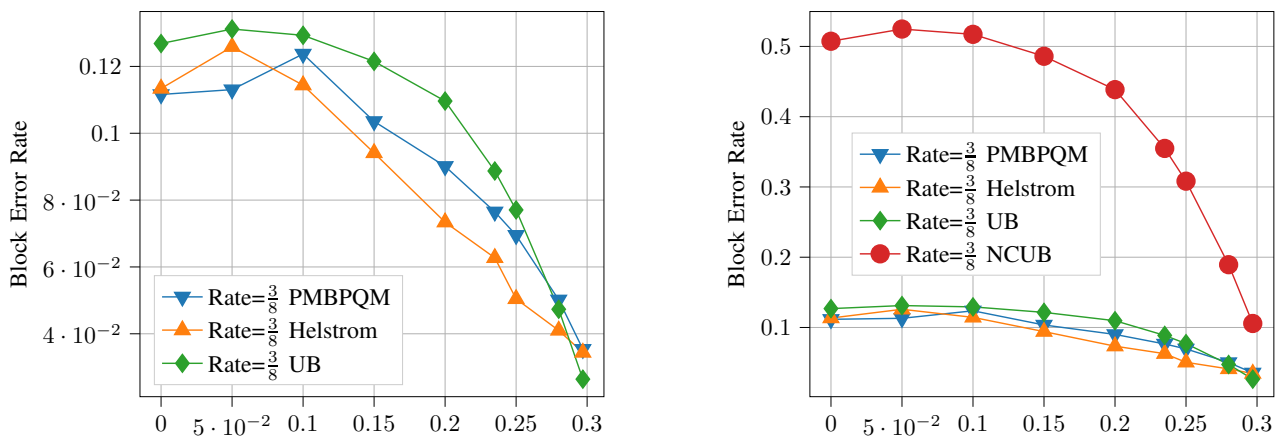

In Fig.~\ref{length8-bler_d=0.1_pmbpqm_helstrom_rate_3/8_ub_ncub}, we compare the block error rate performance of length-8 PM-BPQM polar decoder and successive bitwise Helstrom decoder with union bound (UB) and non-commutative union bound (NCUB) \cite{gao2015quantum} of block error rate obtained from bit error rates for individual channels. Here we consider the code rate to be $\frac{3}{8}$ and the channel parameter $\delta=0.1$.  We observe that the PM-BPQM polar decoder shows almost comparable block error rate performance with the union bound even after the measurement loss while the performance is significantly better compared to the non-commutative union bound.  

\section{Deferred Proofs}\label{deffered proofs}
\begin{IEEEproof}[Proof of Lemma~\ref{lem:bscq-checknode}]
In~\cite{brandsen2022belief} it has been shown that, for check-node combining, the choice of $\ket{v_{0}}$ and $\ket{v_{1}}$ as the paired measurement eigenvectors maximize the mutual information of the post-measurement state with the remainder of the factor graph.
From \cite[Lemma~9]{brandsen2022belief}, we observe that, for qubit-BSCQ channels $W$ and $W'$, the derived channel $W\cnop W'$ is symmetric with symmetry operator $\sigma_{x}\otimes \mathbb{I}_{2}$.
Then, the unitary $C$ is chosen to map the Helstrom matrix ($[W\cnop W'](0)-[W\cnop W'](1)$) eigenvectors onto the computational basis with the first qubit determining the sign of the eigenvalue and the second qubit determining its magnitude.
This implies that its rows, from top to bottom, are given by $\ket{v_{0}}$, $\ket{v_{1}}$, $\sigma_{x}\otimes \mathbb{I}_{2} \ket{v_{0}}$ and $
\sigma_{x}\otimes \mathbb{I}_{2} \ket{v_{1}}$.
By design, it follows that the state after measuring the second qubit is given by $\Pi_0 \tau \Pi_0 + \Pi_1 \tau \Pi_1$ and satisfies \eqref{bscq checknode measurement} with $\{(\delta_{j}^{\cnop},\gamma_{j}^{\cnop})\}_{j=0}^{1}$ given by \eqref{d-checknode} and \eqref{g-checknode} respectively.
\end{IEEEproof}

\begin{IEEEproof}[Proof of Lemma~\ref{bscq-bitnode}]
    From~\cite[Lemma~9]{brandsen2022belief}, we note that, for qubit-BSCQ channels $W$ and $W'$, the derived channel $W\vnop W'$ is symmetric with symmetry operator $\sigma_{x}\otimes \sigma_{x}$.
    Then, the unitary $V(\delta,\gamma,\delta',\gamma')$ is chosen to map the Helstrom matrix ($[W\vnop W'](0)- [W\vnop W'](1)$) eigenvectors onto the computational basis with the first qubit determining the sign of the eigenvalue and the second qubit determining its magnitude.
    This implies that its rows, from top to bottom, are given by $\ket{v_{0}'}$, $\ket{v_{1}'}$, $\sigma_{x}\otimes \sigma_{x} \ket{v_{0}'}$ and $\sigma_{x}\otimes \sigma_{x} \ket{v_{1}'}$.
    By design, it follows that $\Pi_0 \tau' \Pi_0 + \Pi_1 \tau' \Pi_1$ satisfies \eqref{bscq bitnode measurement} with~$\{(\delta_{j}^{\vnop},\gamma_{j}^{\vnop})\}_{j=0}^{1}$ given by \eqref{d-bitnode} and \eqref{g-bitnode}, respectively.
\end{IEEEproof}

\end{appendices}

\end{document}